\documentclass[journal=jacsat,manuscript=article]{achemso}

\usepackage[version=3]{mhchem} 
\usepackage{float}
\usepackage{etoolbox}
\usepackage[utf8]{inputenc}
\usepackage[T1]{fontenc} 
\usepackage{graphicx}
\usepackage[sort&compress]{natbib}
\usepackage{hyperref}
\hypersetup{colorlinks = true, linkcolor = blue, filecolor = magenta, urlcolor=cyan, citecolor=blue, pdfpagemode=FullScreen}
\usepackage{adjustbox}
\usepackage{tabularx}
\usepackage{bm}
\usepackage{mathptmx}
\usepackage{amsmath}
\usepackage{xcolor}
\usepackage[normalem]{ulem}

\author{Komal Sharma}
\affiliation{Department of Physics, Indian Institute of Science, Bangalore, 560012, India}
\author{Riya Dutta}
\affiliation{Department of Physics, Indian Institute of Science, Bangalore, 560012, India}
\altaffiliation{Institute of Physics, University of Rostock, Albert-Einstein-Straße 23, Rostock, 18059, Germany}
\author{Prathmesh Deshmukh}
\affiliation{Department of Physics, City College of New York, 160 Convent Ave., New York, 10031, New York, USA}
\author{Vinod M.Menon}
\affiliation{Department of Physics, City College of New York, 160 Convent Ave., New York, 10031, New York, USA}
\author{Jaydeep K. Basu}
\affiliation{Department of Physics, Indian Institute of Science, Bangalore, 560012, India}
\email{basu@iisc.ac.in}

\title{Cavity-Enhanced Activation of Radiatively Suppressed Light-Hole Exciton Emission in Colloidal Nanoplatelets}


\begin{document}

\begin{abstract}
 Light-hole (LH) excitons provide access to well-defined polarization and spin degrees-of-freedom that are central to quantum photonics and chiral light-matter interactions. Achieving LH emission is challenging because LH states are energetically unfavoured and typically relax non-radiatively. Existing strategies to access LH excitons rely on modifying the electronic band structure through strain, shape anisotropy, or piezoelectric fields, approaches that are material-specific and offer limited post-synthesis tunability. Here we demonstrate an all-photonic route to activate LH exciton emission in colloidal CdSe/CdS nanoplatelets (NPLs) using a distributed Bragg reflector (DBR) cavity, without altering the underlying band structure. In absence of a cavity mode, the system exhibits amplified spontaneous emission from heavy-hole (HH) states without detectable LH emission at low excitation powers. By spectrally matching a cavity resonance to the LH exciton, cavity-coupled LH emission emerges at significantly lower excitation powers. Temperature-dependent spectroscopy reveals reversible switching between LH- and HH-coupled emission through exciton–cavity detuning, while polarization-resolved and spectrally resolved time-resolved photoluminescence measurements provide independent evidence distinguishing the cavity-coupled LH and HH emission channels. These findings establish cavity engineering as a general materials-level approach for accessing radiatively suppressed optical states.

\end{abstract}

Light-hole (LH) excitons have emerged as a subject of significant interest in nanostructured semiconductors owing to their unique dipole orientation and spin configuration. \cite{yuan2018uniaxial,xiang2021coupled, luo2015supercoupling,bataev2022heavy} In contrast to heavy-hole (HH) excitons, which exhibit in-plane dipole moments and total angular momentum J=3/2, LH excitons possess mixed dipole character and J=1/2. \cite{scott2017directed,ithurria2011colloidal,xiang2020electron,slavcheva2010optical} The mixed dipole character of the LH excitons (contributions from $p_x$, $p_y$, and $p_z$) allows optical transitions with both the in-plane and out-of-plane polarization components, thus opening polarization channels inaccessible to HH excitons.\cite{cassette2012colloidal,vezzoli2015exciton}  This versatility makes LH excitons relevant for polarization-resolved emission, spin–photon interfaces, and the development of directional single-photon sources. \cite{zhang2015single,flor2023wavelength,holewa2020optical,filippov2016strongly,gawelczyk2017exciton,ding2025nanomaterials}

In colloidal systems, particularly quasi-two-dimensional CdSe NPLs, the radiative recombination of LH excitons is strongly suppressed. This suppression arises from pronounced quantum confinement along the growth (z) axis and dielectric screening effects, which collectively induce a large HH–LH energy splitting ($\sim$120–160 meV). \cite{shornikova2018addressing,tessier2012spectroscopy,kunneman2014nature} Since LH excitons are energetically unfavored, electrons preferentially relax into HH states upon recombination. Furthermore, the oscillator strength of LH transitions is intrinsically weaker compared to HH transitions, further reducing their radiative contribution. \cite{bataev2022heavy} In addition, the dipole orientation associated with LH excitons couples less efficiently to conventional excitation and collection geometries. Together, these energetic, kinetic, and oscillator-strength considerations explain why LH states are consistently observed in absorption spectra, but rarely contribute to emission under standard experimental conditions.

Epitaxial quantum dots (QDs) have achieved access to LH excitons through strain engineering \cite{huo2014light}, shape anisotropy \cite{jeannin2017light}, or piezoelectric tuning, where band mixing redistributes oscillator strength between HH and LH states \cite{collins1987mixing}. However, such approaches rely on lattice strain or growth-induced asymmetry, limiting post-growth tunability and scalability. In contrast, colloidal QDs and NPLs offer distinct advantages: can be synthesized with atomic-level control over thickness \cite{tessier2012spectroscopy,neeleshwar2005size,christodoulou2018chloride,yu2020optical}, composition \cite{wang2021colloidal,sharma2019two}, and dielectric environment \cite{bertrand2016shape,schlosser2020cds}, enabling systematic tuning of the HH–LH splitting without epitaxial constraints. Their solution processability allows integration with a wide range of photonic structures, such as distributed Bragg reflector (DBR) cavities or dielectric metasurfaces \cite{zhang2016low,morshed2024room,zhao2024efficient,yadav2020room,sharma2024room,wang2024strong,flatten2016strong}, while maintaining strong excitonic confinement and narrow emission linewidths. To date, however, nearly all approaches for accessing LH excitons have focused on engineering the electronic band structure of the material itself. Whether weakly radiative LH states can instead be activated solely through engineering of the surrounding photonic environment remains largely unexplored.

In this context, optical cavities provide a powerful route to manipulate exciton radiative pathways by tailoring the photonic density of states \cite{reithmaier2004strong}. Conventional Fabry–Perot cavities employing metallic mirrors suffer from significant ohmic absorption, low quality factors, and broad spectral response, which limit selective enhancement of weak or energetically unfavored excitonic transitions and can even induce exciton quenching \cite{pfeifer2021achievements}. In contrast, DBR cavities offer a low-loss, high-Q environment with narrow spectral selectivity and minimal absorption, enabling efficient and controlled coupling to targeted excitonic states \cite{liu2015strong}. Such properties make DBR cavities particularly suited for activating radiatively suppressed LH exciton emission.

Among colloidal systems, CdSe/CdS NPLs are particularly promising for investigating LH excitons. \cite{mahler2012core,prudnikau2013cdse,kelestemur2016platelet,rossinelli2017high} Their quasi-two-dimensional electronic structure produces strong and spectrally well-resolved LH absorption bands, reflecting efficient coupling of the optical field to the out-of-plane dipole component. The combination of large oscillator strength, ultranarrow linewidths, and high LH absorbance makes NPLs uniquely suited for photonic activation of otherwise radiatively suppressed LH exciton transitions. \cite{dutta20222d,achtstein2018impact,saidzhonov2019ultrathin}

Here, we experimentally demonstrate a purely photonic route to activate LH exciton emission in colloidal CdSe/CdS NPLs. By coupling the NPL ensemble to a dielectric cavity mode supported by a DBR structure, we demonstrate that resonant field localization can overcome the intrinsic suppression of LH radiative recombination and enable radiative recombination from the LH state. EM simulations reveal at certain NPL layer thicknesses maximize field confinement at the LH energy, while temperature-dependent spectroscopy confirms the photonic origin of the emergent LH emission. Polarization-resolved and spectrally resolved time-resolved photoluminescence measurements further distinguish the cavity-coupled LH and HH emission channels, providing complementary evidence for the assignment of the activated LH emission. This study establishes colloidal NPLs as an ideal testbed for photonic control of exciton states and demonstrates an all-optical mechanism for accessing weakly radiative excitonic transitions in low-dimensional semiconductors.

\begin{figure}[t]
    \centering
    \includegraphics[width=0.9\textwidth]{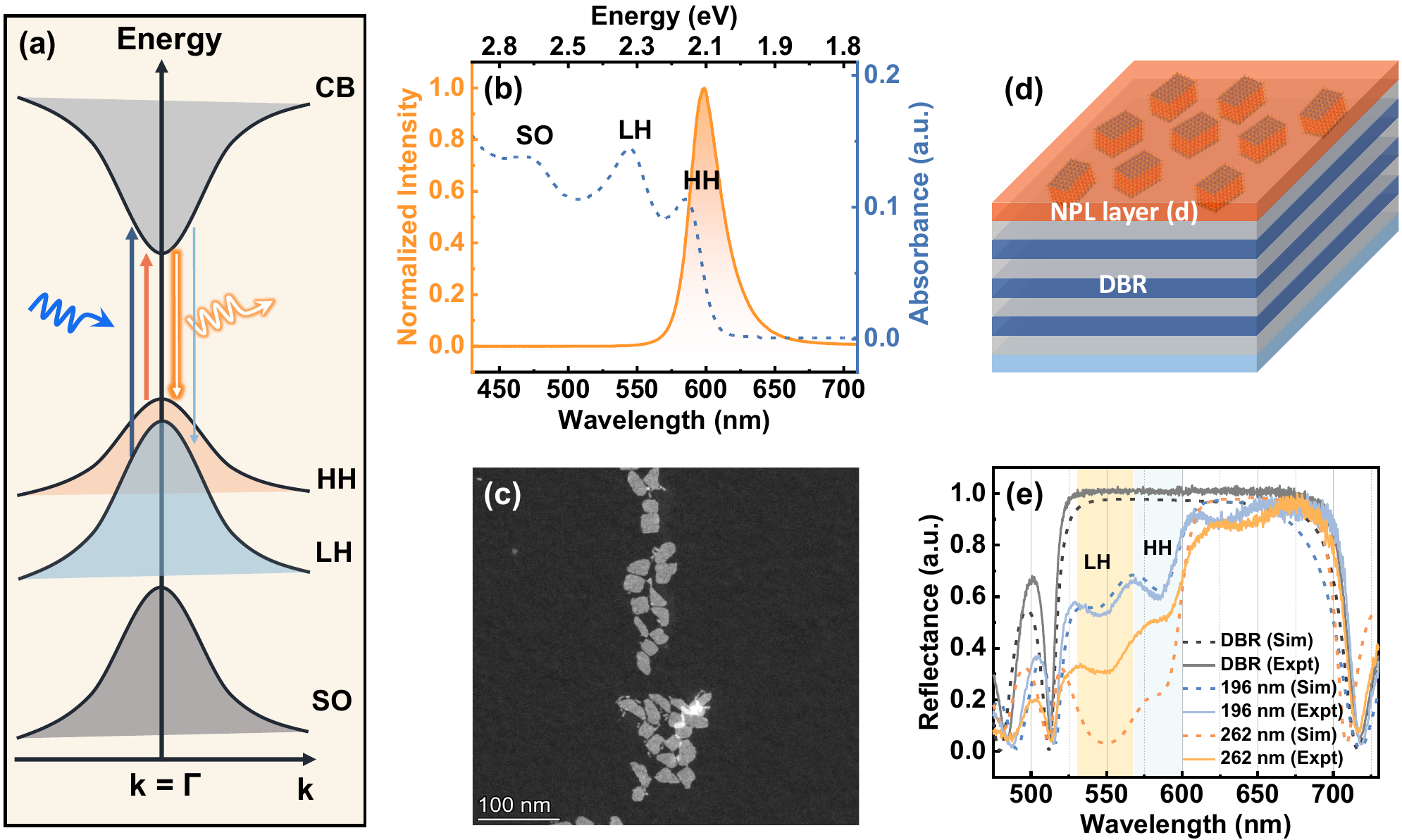}
    \caption{\textbf{(a)} Schematic band diagram of a nanoplatelet (NPL), showing allowed optical transitions from the conduction band (CB) to the heavy-hole (HH), light-hole (LH), and split-off (SO) valence bands at the Brillouin-zone center. While both HH and LH transitions are optically allowed, radiative emission from the LH state is strongly suppressed due to its higher energy and weaker oscillator strength. \textbf{(b)} Normalized photoluminescence (PL) and absorbance spectra of CdSe/CdS core/shell NPLs, showing strong absorption signatures for both HH and LH transitions, but emission exclusively from the HH exciton. \textbf{(c)} Transmission electron microscopy image of NPLs with a scale bar of 100 nm. \textbf{(d)} Schematic of the optical platform: a layer of NPLs deposited on a DBR, forming an open photonic structure that supports interference effects. \textbf{(e)} Simulated and experimental reflectance spectra from DBR substrates with two representative NPL layer thicknesses (196 nm, 262 nm). The reflectance modulation near the HH and LH energies reflects interference effects and sensitivity to the dispersive response of the NPL layer.}
    \label{fig_NPL_DBR}
\end{figure}

To establish the challenge of activating LH exciton emission in colloidal systems, we first examine the intrinsic optical properties of CdSe/CdS NPLs. As illustrated schematically in Fig. \ref{fig_NPL_DBR}a, these quasi-two-dimensional nano-crystals support multiple valence-band transitions involving heavy-hole (HH), light-hole (LH), and split-off (SO) states at the Brillouin-zone center. Experimentally, both HH and LH excitons are clearly resolved in the absorbance spectrum as shown in Fig. \ref{fig_NPL_DBR}b, with an LH-to-HH absorbance ratio of 1.36. In contrast, the steady-state photoluminescence is dominated exclusively by emission from the HH exciton. This behavior reflects the large HH–LH energy splitting in NPLs, which promotes rapid relaxation of photo-excited carriers into the lower-energy HH state. \cite{ithurria2011colloidal}

The structural quality and morphology of the CdSe/CdS NPLs are confirmed by transmission electron microscopy (TEM) presented in Fig. \ref{fig_NPL_DBR}c, which reveals quasi-two-dimensional platelets with well-defined lateral dimensions. Additional high-resolution TEM, HAADF-STEM imaging, and STEM-EDS elemental mapping confirming the crystallinity and elemental composition of the synthesized CdSe/CdS NPLs are provided in Supplementary Fig. S1. Motivated by the strong LH absorption yet negligible LH emission, we next explore whether photonic structures can be used to modify the optical response of the LH exciton. 

To this end, we consider a model system in which an ensemble of NPLs is deposited on top of a DBR, as schematically illustrated in Fig.\ref{fig_NPL_DBR}d. The DBR consists of 10.5 pairs of alternating $SiO_2$ and $TiO_2$ layers terminated with $SiO_2$, forming a vertical slab geometry that supports standing-wave interference. To model the optical response of the NPL–DBR structure, wavelength-dependent complex refractive index values $(n,k)$ for the NPL layer were extracted from UV–Vis absorption measurements and used as input parameters for the transfer matrix method (TMM) simulations (Fig. S2). By varying the thickness of the NPL layer, the optical path length within the structure can be tuned, leading to thickness-dependent redistribution of the EM field and selective field overlap with excitonic resonances. \cite{tischler2006critically} 

Figure \ref{fig_NPL_DBR}e compares simulated and experimental reflectance spectra for DBR substrates coated with NPL layers of two representative thicknesses. For a layer thickness of $d_{\mathrm{NPL}} = 196$ nm, the reflectance modulation at the LH resonance relative to the HH resonance is modest, with an LH/HH contrast of 1.15. In contrast, for a thickness close to the $\lambda/4n$ condition ($d_{\mathrm{NPL}} = 262$ nm; AFM shown in Fig. S3), the LH-related modulation is enhanced, exceeding that of the HH resonance with an effective LH/HH contrast of $\sim$1.38. Despite this pronounced thickness-dependent redistribution of the EM field and the enhanced optical sensitivity at the LH resonance, the emission remains dominated by the HH exciton. As shown in Fig. S4, no detectable LH emission is observed. This demonstrates that interference-induced field enhancement in an open NPL–DBR structure, while sufficient to modify absorption and reflectance, is insufficient to promote radiative recombination from the energetically unfavored LH state. These observations motivate the transition to a closed DBR–NPL–DBR cavity geometry, where stronger spectral confinement and controlled optical feedback can be used to selectively activate LH exciton emission.
For comparison, TMM simulations of a metallic Fabry–Perot cavity (Au–NPL–Au) reveal reduced reflectance $(\sim68\%)$ and absorption-limited confinement in the visible spectral range relevant to LH excitons (Fig. S5), underscoring the advantage of low-loss dielectric DBR cavities for selective light–matter coupling.

\begin{figure}[t]
    \centering
    \includegraphics[width=\textwidth]{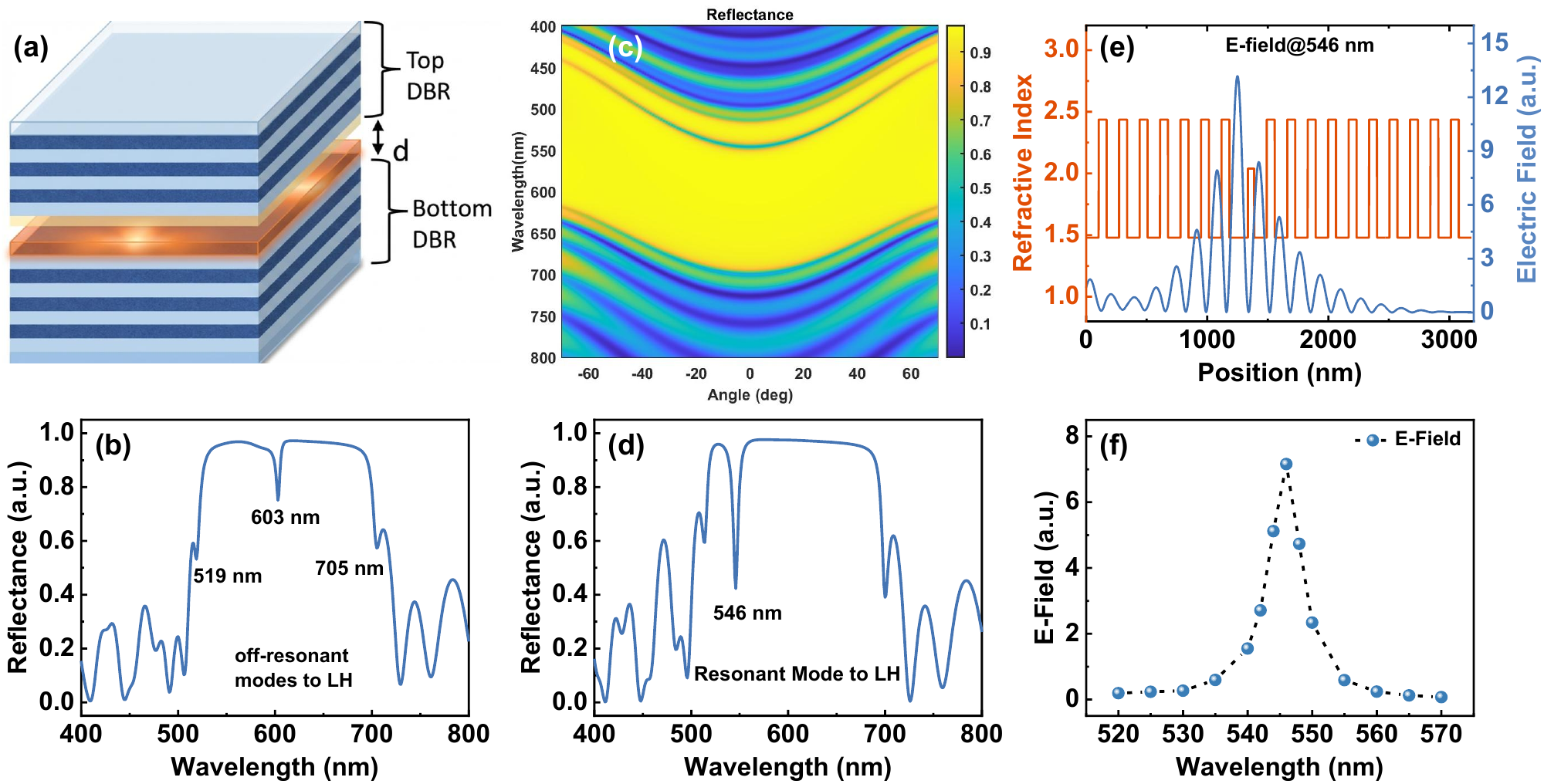}
    \caption{\textbf{(a)} Illustration of the proposed DBR–NPL–DBR cavity structure, designed to spectrally align a cavity mode with the LH exciton. \textbf{(b)} Simulated normal-incidence reflectance spectrum for an initial cavity configuration with an NPL thickness of 262 nm, showing cavity modes at 519, 603, and 705 nm, none of which overlap with the LH exciton. \textbf{(c)} Simulated angle-resolved reflectance spectrum of the optimized cavity structure, revealing a dispersive cavity mode centered near 546 nm that coincides spectrally with the LH exciton. \textbf{(d)} Reflectance profile at normal incidence (0°) confirming a sharp dip at 546 nm, corresponding to the designed resonant cavity. \textbf{(e)} Simulated refractive index profile (orange) and electric field intensity distribution (blue) at 546 nm, showing strong localization of the optical field at the spatial position of the NPL layer embedded within the cavity. \textbf{(f)} Spectral dependence of the electric-field intensity evaluated at the NPL plane, exhibiting a maximum at the LH exciton wavelength.} 
    \label{E-field calc}
\end{figure}

To overcome the limitations of the NPL-DBR structures discussed in Fig. \ref{fig_NPL_DBR}, we designed a vertical DBR–NPL–DBR cavity, as illustrated in Fig. \ref{E-field calc}a, to achieve resonant coupling with the LH exciton. The initial configuration employed the NPL layer thickness of 262 nm, identified earlier as optimal for LH absorption in the open structure, but when this layer was sandwiched between two DBRs, the resulting photonic modes appeared at 519, 605, and 705 nm (Fig. \ref{E-field calc}b), none of which overlapped with the LH transition. We therefore re-optimized the NPL layer thickness to spectrally align a cavity mode with the LH exciton. The simulated angle-resolved reflectance map for this optimized geometry with NPL layer thickness of 65 nm exhibits a well-defined mode centered at 546 nm (2.27 eV), coinciding with the LH absorption band at room temperature as shown in Fig. \ref{E-field calc}c. The normal-incidence reflectance profile depicted in Fig. \ref{E-field calc}d shows a sharp resonance dip at this wavelength, corresponding to a calculated quality factor (Q-factor) of $\simeq182$.

Further insight into the light–matter interaction within the cavity is obtained from the spatial and spectral distribution of the optical field. The simulated electric-field intensity profile at 546 nm, displayed in Fig. \ref{E-field calc}e, shows that the cavity mode exhibits a field antinode at the spatial position of the NPL layer, confirming optimal spatial overlap between the resonant optical mode and the excitonic medium. Such spatial overlap is a necessary condition for efficient coupling, ensuring that excitonic dipoles experience a maximal optical field. Importantly, this field localization alone does not determine radiative recombination rates but rather establishes the geometrical condition for effective interaction.
The spectral selectivity of this interaction is highlighted in Fig.~\ref{E-field calc}f, which presents the wavelength-dependent electric-field intensity evaluated at the NPL position. A pronounced maximum is observed at 546 nm, demonstrating that the cavity provides frequency-selective field enhancement precisely at the LH exciton energy. Together, the spectral resonance, spatial field overlap, and selective field enhancement establish a photonic environment conducive to activating radiative pathways associated with the LH exciton, which are otherwise strongly suppressed under free-space conditions.

\begin{figure}[t]
    \centering
    \includegraphics[width=\textwidth]{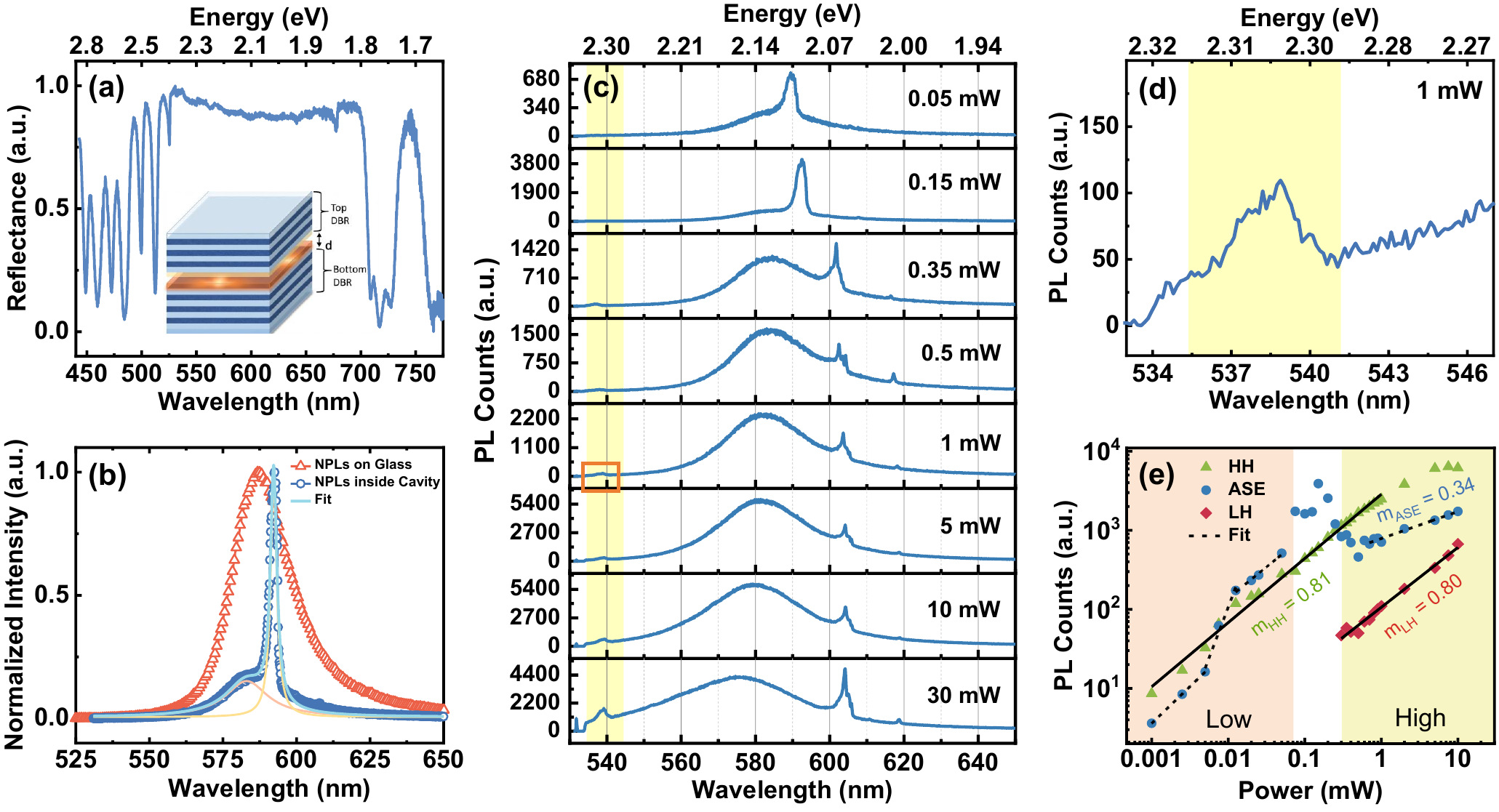}
    \caption{\textbf{(a)} White light reflectance spectrum from the vertical DBR–NPL–DBR cavity geometry, exhibiting no resonant mode to LH. \textbf{(b)} Normalized PL spectra of CdSe/CdS NPLs on glass (red) and embedded inside the DBR cavity (blue), showing significant spectral narrowing and amplification inside the cavity. \textbf{(c)} Power-dependent PL spectra measured from the cavity, showing the progressive emergence of ASE followed by a distinct high-energy peak at the LH exciton energy (shaded). The LH emission is absent at low powers and becomes prominent only beyond a threshold pump fluence. \textbf{(d)} Zoomed-in view of the high-energy spectral region at 1 mW excitation, confirming a weak but discernible LH emission peak. \textbf{(e)} Integrated PL intensity as a function of pump power for the HH, ASE, and LH components, plotted on a log–log scale.} \label{fig_NPL_ASE_Power}
\end{figure}

While the simulations in Fig. \ref{E-field calc} establish the cavity geometry required for spectral resonance with the LH exciton, practical sample preparation inevitably introduces small deviations in layer thickness and effective refractive index that can detune the cavity mode from the target wavelength. To assess the emission dynamics under such non-ideal conditions, we first investigate a DBR–NPL–DBR cavity in which no photonic resonance coincides with the LH exciton energy. Figure \ref{fig_NPL_ASE_Power}a shows the white-light reflectance spectrum of this detuned cavity, confirming the absence of a cavity mode at the LH position. Despite the lack of spectral resonance, the cavity substantially modifies the emission characteristics under continuous-wave (CW) excitation at 532 nm. As shown in Fig. \ref{fig_NPL_ASE_Power}b, the broad PL spectrum of CdSe/CdS NPLs on glass (linewidth $\sim$ 22 nm (80 meV)) becomes strongly intensified and spectrally narrowed to 2.6 nm (9.24 meV) inside the cavity, consistent with enhanced optical feedback and cavity-mediated modification of spontaneous emission processes. \cite{duan2025continuous}

At low excitation powers, the emission remains dominated by the HH exciton. With increasing pump power (Fig. \ref{fig_NPL_ASE_Power}c), a pronounced ASE feature emerges on the low-energy side of the HH transition with a threshold of 0.003 mW. This regime is characterized by rapid linewidth narrowing and a superlinear increase in intensity with a slope of 2.29, consistent with cavity-assisted gain processes. Time-resolved PL (TRPL) measurements reveal a reduction of the average exciton lifetime from 22 ns on a DBR to 8 ns in the non-resonant cavity, corresponding to a modest enhancement of the radiative rate that supports ASE (Fig. S6 and Table S1). This indicates that even in the absence of spectral resonance, the cavity modifies the local photonic environment sufficiently to enhance spontaneous emission rates and facilitate ASE.

At higher excitation powers (1 mW), the ASE signal begins to saturate (slope $\approx$0.34), and an additional emission feature emerges at higher energy near 2.3 eV (540 nm), coinciding with the LH absorption resonance identified in Fig. \ref{fig_NPL_DBR}. A magnified view of this spectral region (Fig. \ref{fig_NPL_ASE_Power}d) confirms the presence of a weak but distinct LH-related emission peak. The integrated intensity of this feature exhibits a sublinear power dependence with a slope of 0.80, comparable to that of the HH exciton (0.81), as summarized in Fig. \ref{fig_NPL_ASE_Power}e. While these results demonstrate that cavity geometry and high excitation power can partially activate the LH exciton, the emission remains weak and power-sensitive in the absence of spectral resonance. This raises a key question: Can the LH transition be efficiently activated at lower excitation levels if the cavity mode is tuned into resonance with it? Temperature-dependent PL measurements on C/S NPLs deposited on glass show no emergence of LH emission across the entire temperature range, establishing that thermal suppression of non-radiative pathways alone does not activate the LH exciton (Fig. S7).

\begin{figure}[t]
    \centering
    \includegraphics[width=\textwidth]{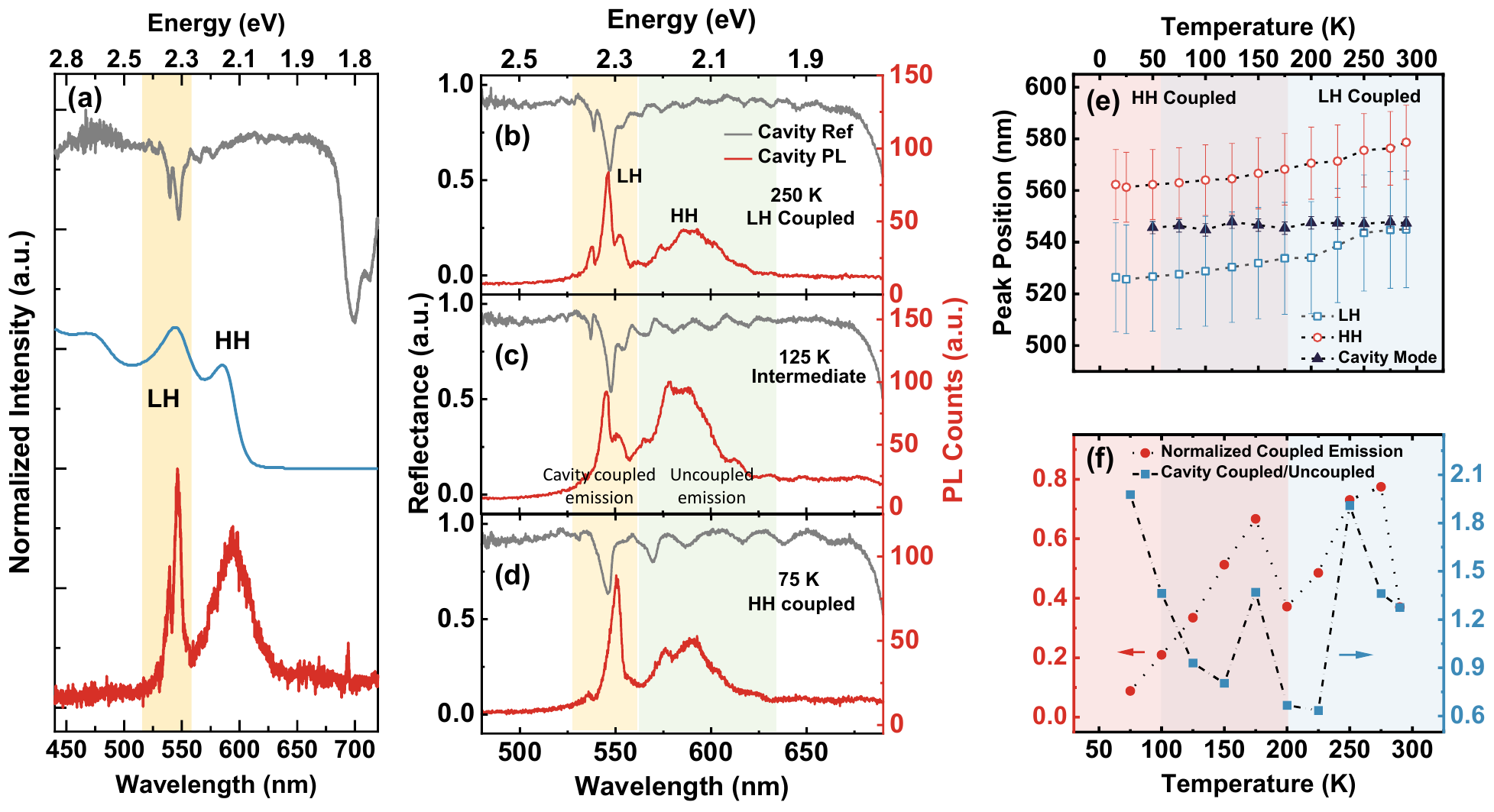}
    \caption{\textbf{(a)} Measured cavity reflectance (gray) at room temperature (297 K) and NPL absorbance (blue), showing that the cavity mode is spectrally aligned with the LH absorption resonance. The corresponding PL spectrum (red) from NPLs coupled to the cavity exhibits strong emission at the LH energy, indicating activation of the weakly radiative LH transition. \textbf{(b-d)} Measured cavity reflectance (gray), and cavity-coupled PL (red) at \textbf{(b)} 250 K, \textbf{(c)} 125 K, and \textbf{(d)} 75 K. \textbf{(e)} Extracted peak positions of the LH, HH, and cavity mode as a function of temperature from reflectance measurements collected on NPL-DBR sample with 262 nm layer thickness. \textbf{(f)} LH emission intensity normalized with respect to the intrinsic (uncoupled) PL of the NPLs corresponding to HH, plotted as a function of temperature. Ratio of cavity coupled emission to uncoupled HH emission with temperature highlighting local maxima corresponding to LH-coupled and HH-coupled regimes.} 
    \label{Cavity Coupled LH}
\end{figure}

To directly probe the role of spectral resonance, we fabricated a cavity whose optical mode is tuned into alignment with the LH exciton absorption band at room temperature. This was achieved by adjusting the NPL layer thickness to 65 nm, as guided by the simulations shown in Fig. \ref{E-field calc}, such that the cavity resonance coincides with the LH transition energy. To isolate the effect of exciton–cavity spectral matching without modifying the photonic geometry, temperature-dependent measurements were employed to systematically tune the excitonic transitions, while the spectral position of the cavity mode remains essentially insensitive to temperature \cite{varshni1967temperature}.

As shown in Fig. \ref{Cavity Coupled LH}a, the room-temperature reflectance spectrum exhibits a cavity mode at 548 nm (2.26 eV) with a quality factor of approximately 100, spectrally aligned with the LH absorption peak of the NPLs. Under these conditions, the cavity-coupled PL spectrum displays a pronounced LH emission feature in addition to the intrinsic HH emission, indicating efficient resonant coupling between the cavity mode and the LH exciton. Power-dependent measurements further show that the cavity-coupled LH and HH emission intensities increase with nearly identical sublinear power-law exponents ($m_{\mathrm{LH}} = 0.89$ and $m_{\mathrm{HH}} = 0.87$), without any threshold-like behavior (Fig. S8). This confirms that the resonantly activated LH feature originates from spontaneous excitonic emission rather than ASE or stimulated emission. To further verify the resonant origin of this emission, temperature-dependent reflectance and PL measurements were performed on the cavity-integrated C/S NPL system.

Upon decreasing the temperature, both LH and HH excitonic transitions exhibit a characteristic blueshift due to bandgap renormalization \cite{varshni1967temperature} as shown in Fig. S9. As a consequence, the LH exciton progressively detunes from the cavity mode (Fig. \ref{Cavity Coupled LH}b-d), leading to a suppression of the cavity-activated LH emission. Simultaneously, the HH exciton, which lies at lower energy, moves closer to spectral resonance with the cavity mode, resulting in a crossover from an LH-dominated to an HH-dominated coupling regime at low temperatures. The extracted peak positions of the LH, HH, and cavity mode shown in Fig. \ref{Cavity Coupled LH}e reveal excitonic blueshifts approaching 25 nm (100 meV) across the measured temperature range, while the cavity resonance remains effectively stationary. Temperature-dependent PL and reflectance spectra at additional temperatures, together with PL from NPLs on glass for comparison, are provided in Supplementary Fig. S10 and confirm the absence of LH emission in the intrinsic NPL response without cavity coupling.

The temperature dependence of the cavity-modified emission is summarized in Fig. \ref{Cavity Coupled LH}f. The red circles represent the LH emission intensity from the cavity-coupled system normalized to the intrinsic HH emission of uncoupled NPLs on glass, providing a direct measure of cavity-induced activation of the weakly radiative LH transition. This normalized LH emission exhibits a non-monotonic temperature dependence, reaching a maximum when the LH exciton is spectrally resonant with the cavity mode. The blue squares denote the LH-to-HH intensity ratio extracted from the cavity-coupled PL spectra. Around room temperatures, this ratio approaches a value of approximately 2, indicating efficient cavity-enhanced LH emission. As the temperature is reduced, the ratio decreases due to progressive detuning of the LH exciton from the cavity resonance. Below approximately 150 K, the ratio increases again to a comparable magnitude; however, in this low-temperature regime the enhancement arises from cavity coupling to the HH exciton rather than renewed LH coupling. Together, these trends demonstrate that radiative recombination in the NPL–DBR system is governed by temperature-controlled spectral matching between the cavity mode and excitonic transitions, enabling selective coupling to either LH or HH excitons. As an additional control, a cavity designed with a resonance close to the HH transition exhibits a monotonic increase in the integrated cavity-coupled emission as the HH exciton approaches resonance upon cooling (Fig. S11). This behavior contrasts with the non-monotonic response of the LH-coupled cavity and further confirms that the observed temperature dependence is governed by exciton--cavity detuning.

Importantly, control experiments performed on core-CdSe NPLs embedded in identical cavity geometries show no detectable LH emission under comparable conditions as shown in Fig. S12, confirming that both material engineering and photonic design are essential for activating this weakly radiative transition.

\begin{figure}[t]
    \centering
    \includegraphics[width=0.8\textwidth]{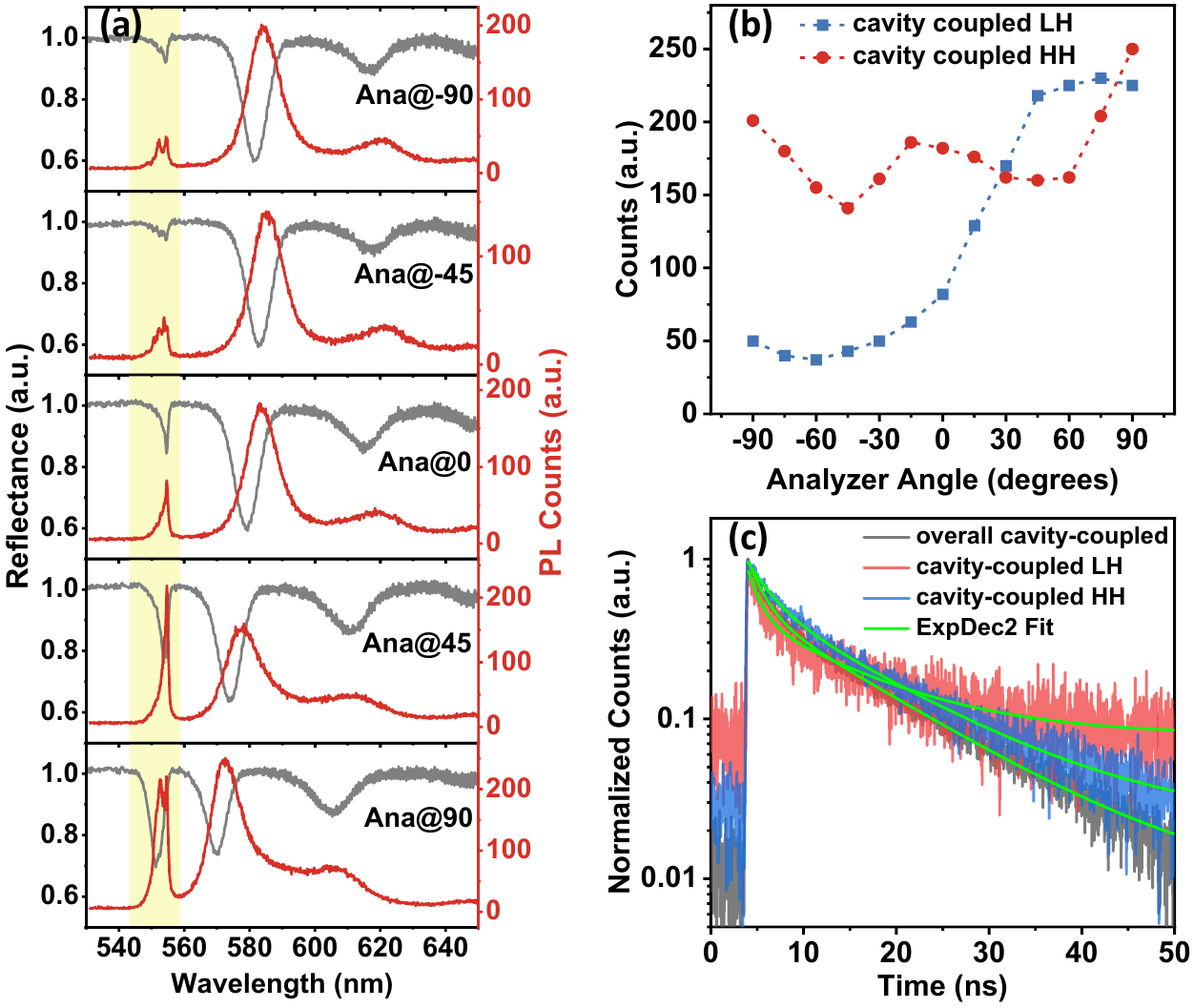}
    \caption{\textbf{(a)} Polarization-resolved cavity reflectance (gray) and cavity-coupled photoluminescence (red) measured for different analyzer angles. The shaded region highlights the cavity-coupled LH emission, while the HH cavity mode is centered near $\sim590$ nm. \textbf{(b)} Peak PL intensity of the cavity-coupled LH ($\sim552$ nm) and cavity-coupled HH ($\sim590$ nm) emission as a function of analyzer angle, showing distinct polarization dependences for the two cavity-coupled emission channels. \textbf{(c)} Normalized time-resolved photoluminescence (TRPL) decay profiles of the overall cavity-coupled emission, cavity-coupled LH emission (550 $\pm$ 5 nm), and cavity-coupled HH emission (590 $\pm$ 5 nm), together with biexponential fits (green dashed lines).}
    \label{Fig_Pol_TRPL}
\end{figure}

Although the spectral position and temperature dependence strongly support the assignment of the high-energy feature to cavity-coupled LH emission, these observations alone do not uniquely establish its excitonic origin. We therefore performed polarization-resolved spectroscopy and spectrally resolved TRPL measurements.

Figure \ref{Fig_Pol_TRPL}(a,b) compares the polarization response of the cavity-coupled LH and cavity-coupled HH emission channels. Both emission features exhibit clear analyzer-angle-dependent modulation, confirming that they retain the polarization characteristics imposed by their respective cavity resonances. Interestingly, despite originating from the same cavity supporting resonances near the LH and HH transitions, the two emission channels exhibit distinctly different polarization dependences, with different analyzer angles corresponding to their respective intensity maxima and minima. Quantitatively, the cavity-coupled LH emission exhibits a larger degree of polarization (DOP = 0.72, PER = 6.2) than the cavity-coupled HH emission (DOP = 0.28, PER = 1.7), indicating that the two transitions interact differently
with the anisotropic cavity modes. The corresponding polarization-dependent reflectance measurements further reveal that both cavity resonances exhibit analyzer-angle-dependent modulation, corroborating the anisotropic optical response of the coupled NPL--cavity system. For comparison, polarization-resolved measurements from an HH-control cavity and from uncoupled NPLs on glass are provided in Fig. S13 and Table S2. The cavity-coupled LH emission exhibits substantially stronger polarization modulation than the HH-control cavity and uncoupled NPL emission, supporting a distinct interaction of the LH-related transition with the cavity mode.

To further distinguish the two emission channels, spectrally resolved TRPL measurements were performed by selectively detecting the cavity-coupled LH (550 $\pm$ 5 nm) and HH (590 $\pm$ 5 nm) emission windows [Fig. \ref{Fig_Pol_TRPL}(c)]. Both decay profiles are well described by a bi-exponential model; however, the cavity-coupled LH emission exhibits a faster dominant decay component than the cavity-coupled HH emission, consistent with stronger radiative coupling at the LH resonance. While neither polarization nor TRPL measurements alone provide an unambiguous identification of the excitonic origin, together with the spectral, temperature-dependent, power-dependent, and control cavity measurements presented in the Supporting Information, these independent observables consistently suggest the assignment of the high-energy emission to cavity-coupled LH excitons.

\begin{table}
\centering
\begin{tabular}{ c  c  c  c  c  c } 
\hline
\hline
System & $A_1$ & $\tau_1$ (ns) & $A_2$ & $\tau_2$ (ns) & $\tau_{avg}$ (ns) \\
\hline
\hline
overall cavity-coupled 
& $0.890$ & $1.69$ & $0.110$ & $12.16$ & $6.61$ \\
 & $\pm 0.004$ & $\pm 0.03$ & $\pm 0.004$ & $\pm 0.15$ & $\pm 0.14$\\
\hline
cavity-coupled LH 
& $0.948$ & $1.30$ & $0.052$ & $11.14$ & $4.43$ \\
& $\pm 0.012$ & $\pm 0.10$ & $\pm 0.012$ & $\pm 0.62$ & $\pm 0.60$ \\
\hline
cavity-coupled HH 
& $0.743$ & $2.52$ & $0.257$ & $12.51$ & $8.84$ \\
& $\pm 0.011$ & $\pm 0.10$ & $\pm 0.011$ & $\pm 0.37$ & $\pm 0.33$ \\
\hline
\end{tabular}
\caption{\textbf{Summary of lifetime components, normalized weight factors, and average lifetimes for cavity-coupled LH and HH emission.}}
\label{tab:TRPL_Cavity}
\end{table}


In summary, we have demonstrated that a dielectric DBR cavity can activate radiatively suppressed light-hole exciton emission in colloidal CdSe/CdS NPLs without modifying their underlying electronic band structure. By spectrally matching the cavity resonance to the LH transition, we selectively enhanced an otherwise inaccessible radiative pathway, while temperature-dependent spectroscopy enabled reversible switching between LH- and HH-coupled emission regimes. Polarization-resolved and spectrally resolved time-resolved measurements further distinguished the cavity-coupled LH and HH emission channels, providing complementary evidence for the activated LH emission. These results establish photonic density-of-states engineering as a versatile strategy for accessing weakly radiative excitonic states and open new opportunities for polarization-controlled quantum light sources, spin-photon interfaces, and cavity-engineered colloidal nanophotonics.

\section*{Experimental Methods}

\subsection*{Materials}
Cadmium nitrate tetrahydrate $(Cd(NO_{3})_{2}.4H_{2}O)$, Cadmium Acetate Dihydrate, Sodium Hydroxide (NaOH), Myristic Acid, Selenium (Se), 1-Octadecene (ODE), N-methylformamide (NMF), Ammonium sulfide ($(NH_4)_2S_2$), Oleic Acid (OA), Oleylamine, Acetonitrile, Ethanol, Toluene and Hexane  were obtained from Sigma-Aldrich (Germany) for the synthesis of NPLs.

\subsection*{CdSe Core NPLs}

In a three neck flask, 170 mg of $Cd(myr)_{2}$ and 14 mL of ODE were introduced and the flask was degassed for 1 hr at room temperature following the protocol \cite{ithurria2011continuous}. The mixture was heated to 240\textdegree{C} under nitrogen flow. A solution of 12 mg of Se dispersed in 1 mL of ODE was quickly injected to the flask. One minute later, 120 mg of $Cd(Ac)_{2}.2H_{2}O$ was introduced to the flask. The mixture was heated for 10 min at 250\textdegree{C}. The NPLs were separated from the QDs using selective precipitation. The NPLs were then dispersed in hexane.

\subsection*{Synthesis of CdSe/CdS Core/Shell NPLs}

In order to grow a shell on core NPLs, we follow the procol described in the article \cite{ithurria2012colloidal}. We start with 4 mL solution of core NPLs in hexane and add 1 mL of N-methylformamide (NMF), along with 20 $\mu$L of ammonium sulfide ($(NH_4)_2S_2$), stir the mixture for 1 min mildly and wait for phase transfer from hexane to NMF. The hexane solvent was discarded and again 4 mL of hexane was added, stirred and discarded. To remove the excess $S^{2-}$, 1.5 mL of acetonitrile and 4 mL of toluene was added and the NPLs were precipitated by centrifugation at 6000 rpm for 4 mins. This cleaning step was again repeated by redispersing the precipitates in 2 mL of NMF as it is crucial to prevent the secondary nucleation of CdS in the next step. To this solution of NPLs in 2 mL NMF, 1.75 mL of 0.2 M of cadmium acetate was introduced and stirred for a while. Again, the solution was precipitated with 10 mL of toluene and redispersed in 1 mL of NMF. To this solution, 200 $\mu$L of oleylamine dispersed in 5 mL of hexane was added and mixture was left for phase transfer from NMF to hexane. Finally, the oleylamine-capped CdSe/CdS NPLs were collected from top layer of hexane.

\subsection*{Sample Preparation}

The synthesized NPLs were deposited onto the DBR mirror using a spin-coating technique. The NPL solution (40 mg/mL) was mixed in a PMMA matrix for the deposition and the spin speeds were varied from 1000 to 3000 rpm to achieve the desired layer thicknesses. The DBR mirrors used in this study consisted of 10.5 alternating pairs of $SiO_{2}$ ($n \sim 1.45$) and $TiO_{2}$ ($n \sim 2.25$), with individual layer thicknesses of 104 nm and 64 nm, respectively.

To fabricate the planar DBR–NPL–DBR cavity, a 30 mg/mL solution of CdSe/CdS NPLs in hexane was spin-coated onto the bottom DBR. A second PMMA-coated DBR was then gently placed on top using a pick-and-place approach to form the cavity.

\subsection*{White Light Reflectance and Photo-luminescence Measurements}

Reflectance, photoluminescence (PL), and time-resolved photoluminescence (TRPL) measurements were performed using a WITec confocal system (WITec alpha-300R). White-light reflectance measurements were conducted on the NPL-coated DBR using an LED source with a 50 $\mathrm{\mu m}$ aperture. Steady-state cavity PL measurements (Fig. 3) were performed using a continuous-wave (CW) 532 nm laser, whereas the cavity PL and TRPL measurements presented in Figs. 4 and 5 were carried out using a 405 nm pulsed diode laser (20 MHz repetition rate, pulse width 30 ps) with an excitation power of 20 $\mathrm{\mu W}$ measured before the objective. All measurements were performed in a reflection geometry.

The NPLs were excited from the top using a 20$\times$ / 0.22 NA objective, which was also used to collect the emitted signal. The collected signal was directed to a spectrometer equipped with a 600 grooves/mm grating and detected using a charge-coupled device (CCD) detector. For TRPL measurements, the spectrally selected emission was directed to an avalanche photodiode (APD) through band-pass filters centered at $\sim$550 nm and $\sim$590 nm to isolate the cavity-coupled LH and HH emission channels, respectively. The overall instrument response function (IRF) of the TRPL system was 300 ps.

\begin{acknowledgement}

J.K.B. acknowledges funding from science and engineering research board (SERB), India through grant number CRG/2021/003026, and DST, FIST grant. K.S. would like to thank Ministry of Education, India and Scheme for Promotion of Academic and Research Collaboration (SPARC) program for the fellowship support. K.S. thanks Pinaki Chatterjee for the fruitful discussions. K.S. acknowledges Dr. Binita Tongbram for discussions related to TEM.

\end{acknowledgement}

\section*{Author Contribution}
K.S. synthesized NPLs. K.S., R.D. and P.D. conceptualized the experiments. K.S. performed all the experiments, data analysis and drafted the manuscript. J.K.B. and V.M.M. provided guidance for data analysis. J.K.B. and V.M.M. initiated and supervised the project. All authors contributed to the manuscript writing and discussions.




\newpage
\begin{center}
\section{SUPPORTING INFORMATION}
\end{center}

\setcounter{section}{0}
\renewcommand{\thesection}{S\arabic{section}}

\setcounter{table}{0}
\renewcommand{\tablename}{Tab.}
\renewcommand{\thetable}{S\arabic{table}}
\renewcommand{\theHtable}{S.\arabic{table}}

\setcounter{figure}{0}
\renewcommand{\figurename}{Fig.}
\renewcommand{\thefigure}{S\arabic{figure}}
\renewcommand{\theHfigure}{S.\arabic{figure}}

\setcounter{equation}{0}
\renewcommand{\theequation}{S\arabic{equation}}
\renewcommand{\theHequation}{S.\arabic{equation}}

\section{Structural Characterization of CdSe/CdS NPLs}

\begin{figure}
    \centering
    \includegraphics[width=1\textwidth]{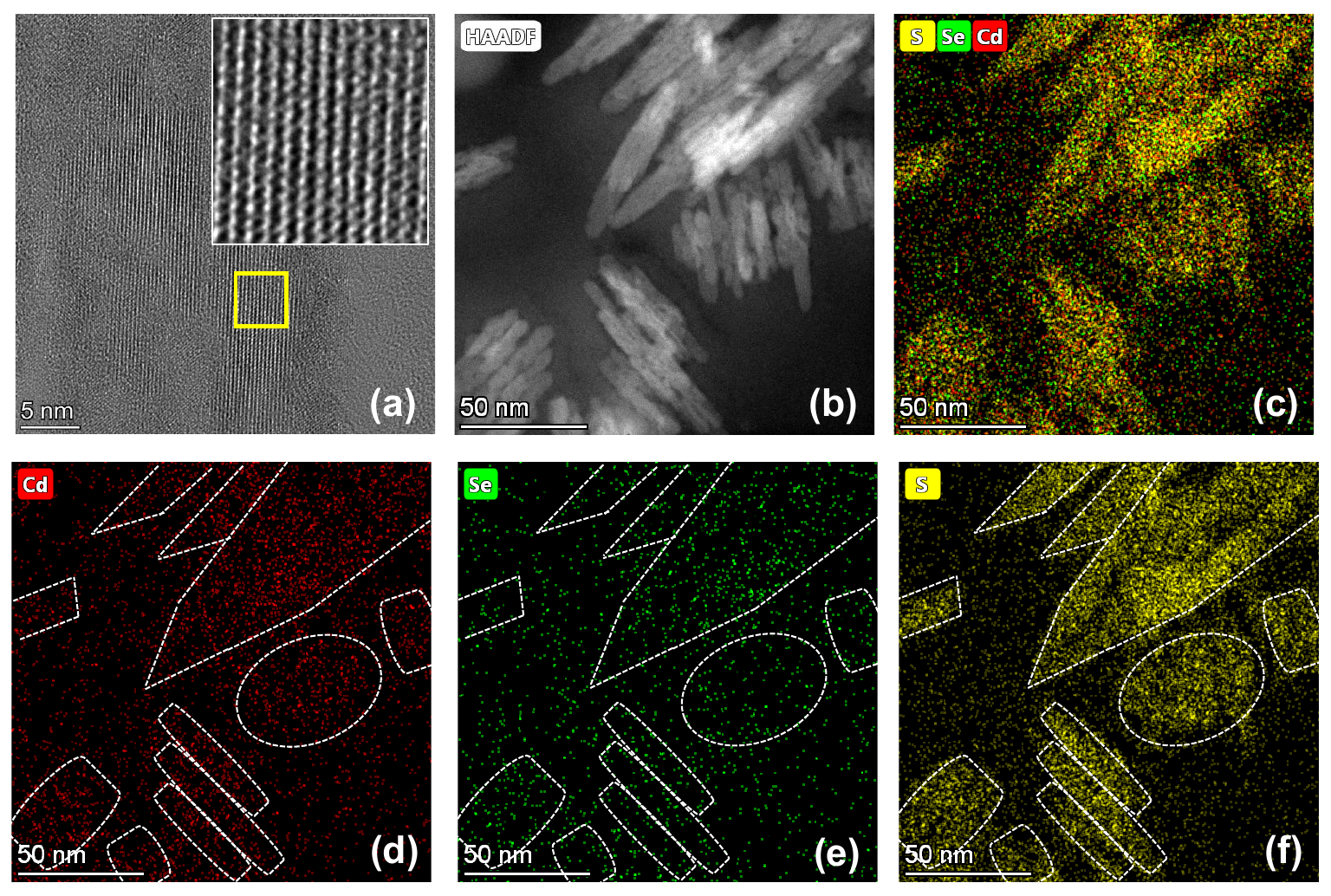}
    \caption{(a) HRTEM image showing well-resolved lattice fringes. The enlarged region marked by the yellow box is shown in the inset; the measured inter-planar spacing (d) is $3.3\pm 0.2$ $\mathring{\mathrm{A}}$. (b) HAADF-STEM image of the synthesized CdSe/CdS NPLs. (c) Composite STEM-EDS elemental map showing the spatial distribution of Cd (red), Se (green), and S (yellow). (d-f) Individual elemental maps corresponding to Cd, Se, and S, respectively, confirming the presence of the constituent elements within the CdSe/CdS NPLs.}
    \label{SI_HAADF_TEM}
    \end{figure}

 High-resolution TEM (HRTEM) images have been included to assess the crystalline quality of the NPLs as shown in Fig. \ref{SI_HAADF_TEM}(a). Well-resolved lattice fringes are observed throughout the platelet, and fast Fourier transform (FFT) analysis yields an interplanar spacing of $3.3\pm 0.2$ $\mathring{\mathrm{A}}$, consistent with the reported $d_{111}$ lattice spacing of zinc-blende CdSe/CdS nanocrystals, confirming the high crystallinity of the synthesized NPLs.\cite{cassette2012colloidal}

To further investigate the composition of the synthesized NPLs, STEM-EDS elemental mapping has been performed. Figure \ref{SI_HAADF_TEM} shows the composite elemental map together with the individual Cd, Se, and S elemental maps, confirming the presence of all constituent elements within the synthesized NPL ensemble. These measurements complement the HAADF-STEM and HRTEM analyses and, together with the characteristic absorption and photoluminescence spectra of the synthesized samples, provide mutually consistent evidence supporting the successful synthesis of CdSe/CdS core/shell NPLs.

\section{Optical Constants of NPLs}

The effective optical constants of the NPL layers were obtained by combining UV--Vis absorption measurements with Kramers--Kronig analysis. \cite{lucarini2005kramers} The complex refractive index is written as
\begin{equation}
N(\omega)=\eta(\omega)+i\kappa(\omega),
\end{equation}
where $\eta(\omega)$ is the real part of the refractive index and $\kappa(\omega)$ is the extinction coefficient.

The extinction spectrum was first obtained from the UV-Vis absorbance measured on a colloidal NPL solution. The solution-phase absorbance captures the intrinsic excitonic absorption profile of the nanoplatelets, including the heavy-hole (HH), light-hole (LH), and split-off (SO) transitions. The measured absorbance spectrum was converted from wavelength to energy and spline-interpolated over the range 1.77--2.92 eV using 10000 equally spaced energy points, as required for the numerical Kramers-Kronig transformation.

The real dispersive contribution to the refractive index was then calculated using the Kramers--Kronig relation for the complex refractive index,

\begin{equation}
    \Delta \eta_{\mathrm{KK}}(\omega)=
    \frac{2}{\pi}\,
    \mathcal{P}
    \int_{0}^{\infty}
    \frac{\omega'\,\kappa(\omega')}
    {\omega'^{2}-\omega^{2}}\,
    d\omega',
\end{equation}

where $\mathcal{P}$ denotes the Cauchy principal value. In the numerical implementation, the principal-value integral was evaluated by excluding the singular point at $\omega'=\omega$ from the discrete summation.

Since the measured spectrum spans a finite energy range, the Kramers--Kronig transformation provides the dispersive refractive-index variation associated with the measured excitonic absorption bands. The final real refractive index was obtained by adding a constant non-resonant background term:

\begin{equation}
    \eta(\omega)=\eta_{\mathrm{bg}}+\Delta \eta_{\mathrm{KK}}(\omega),
\end{equation}

where $\eta_{\mathrm{bg}}=1.9$ was used as the effective background refractive index. This value accounts for the non-resonant optical response of the NPL/PMMA composite film in the visible spectral region. The calculated $\eta(\omega)$ values were subsequently converted to wavelength using $\lambda = 1240/E$, where $E$ is expressed in eV and $\lambda$ in nm.

\begin{figure}
    \centering
    \includegraphics[width=0.8\textwidth]{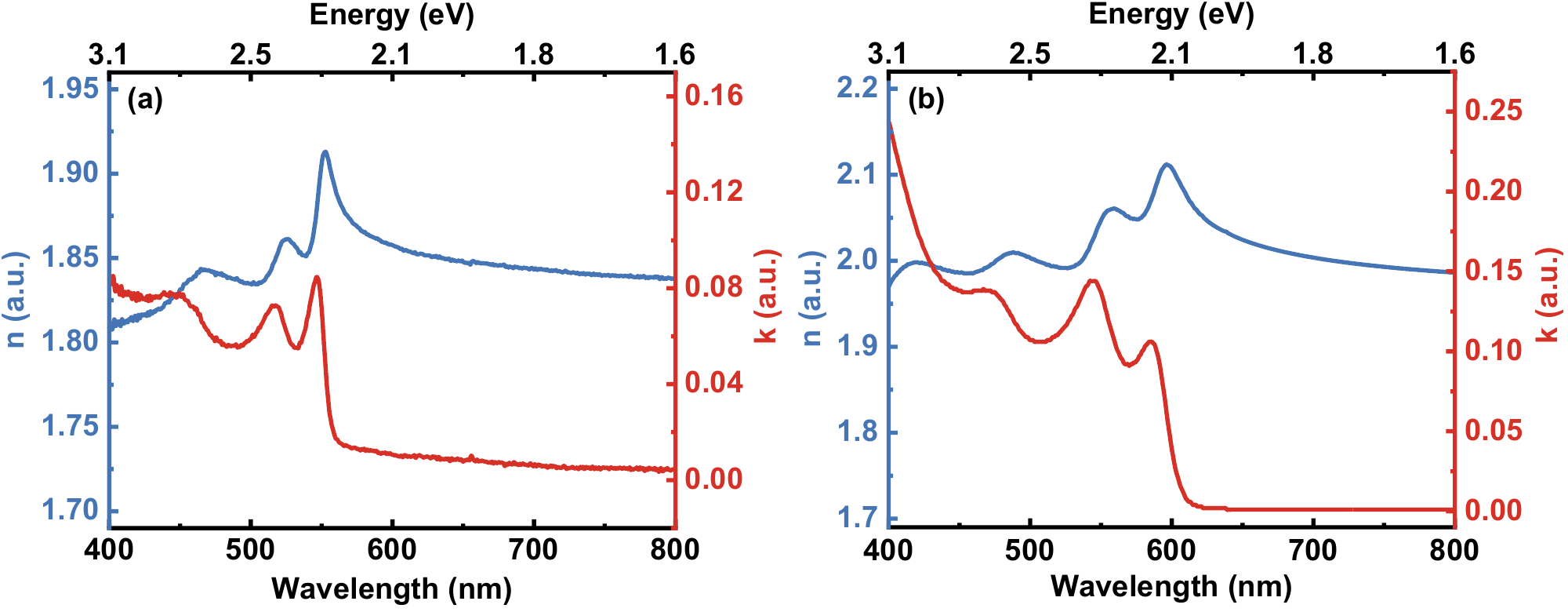}
    \caption{n and k values for \textbf{(a)} core CdSe NPLs and \textbf{(b)} core/shell CdSe/CdS NPLs.} 
    \label{SI_n_K}
\end{figure}

We note that since the UV--Vis measurement was performed on a colloidal NPL solution, the extracted $\kappa(\omega)$ captures the intrinsic excitonic absorption profile of the NPLs, while the background refractive index was chosen to represent the effective non-resonant response of the spin-coated NPL/PMMA film used in the photonic structures. The resulting $\eta(\lambda)$ and $\kappa(\lambda)$ spectra were treated as effective optical constants of the NPL composite layer and were used as input parameters for the transfer-matrix method (TMM) simulations. Optical simulations of the DBR microcavity were performed using the MATLAB-based Reflectance–Absorbance–Transmittance (RAT Catcher) transfer-matrix package developed by Galfsky and Menon. The multilayer stack was modeled using the experimentally measured layer thicknesses and wavelength-dependent refractive indices of the constituent materials. The cavity reflectance spectra were calculated under normal incidence for TM polarization, and the cavity resonance was obtained from the reflectance minimum. The validity of the extracted effective optical constants is supported by the agreement between the simulated and measured reflectance spectra shown in Figure 1e of the main text.

The wavelength-dependent $n$ and $k$ values for C-NPLs and C/S-NPLs are shown in Figs.\ref{SI_n_K}(a) and \ref{SI_n_K}(b), respectively. The stronger extinction feature near the LH absorption resonance in the C/S NPLs reflects their enhanced LH absorption strength.

\section{Morphological Characterization of NPL Films}

\begin{figure}
    \centering
    \includegraphics[width=1\textwidth]{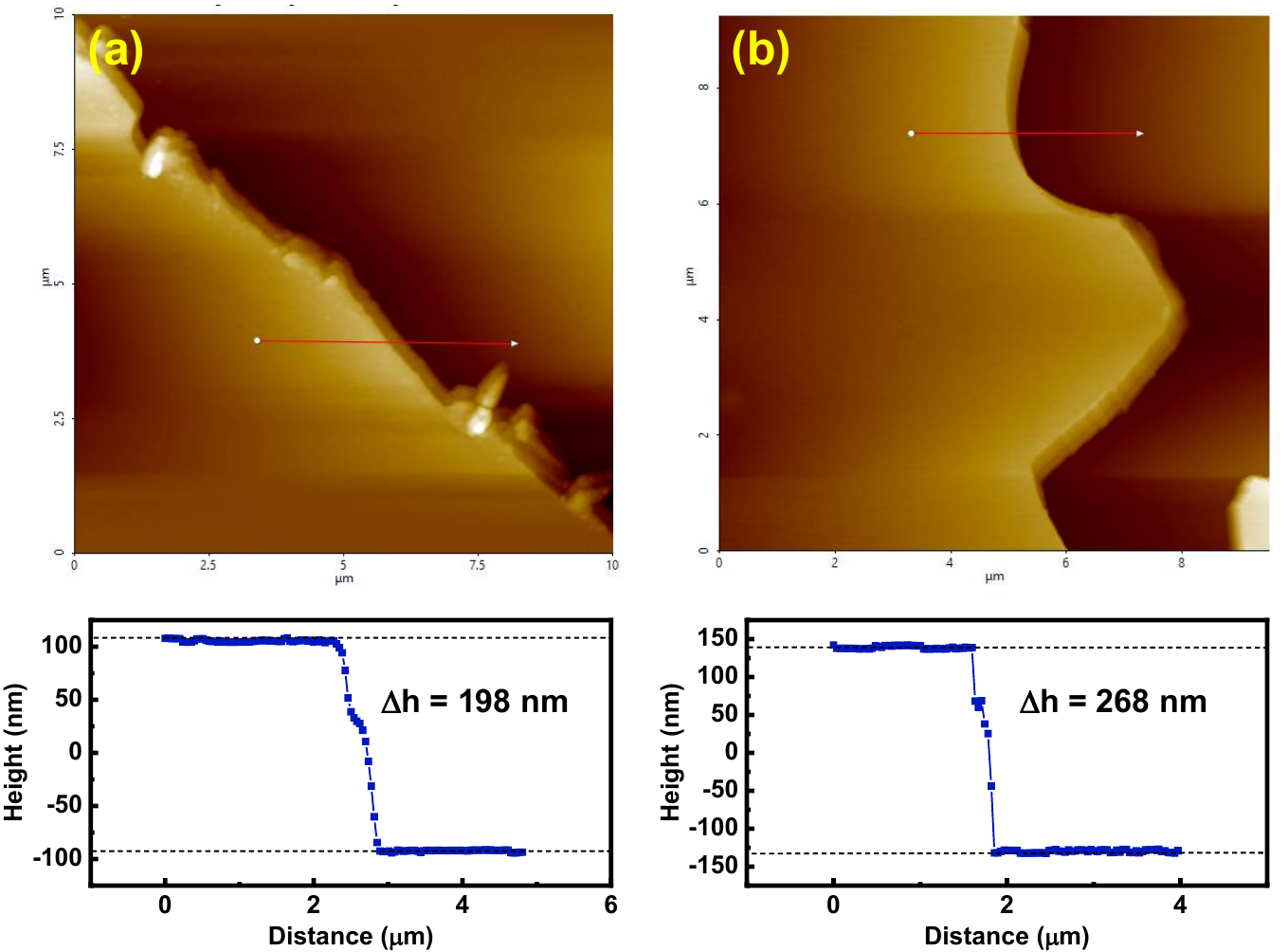}
    \caption{AFM images and corresponding height profiles of C/S NPL films with measured thicknesses of \textbf{(a)} 198 nm and \textbf{(b)} 268 nm, corresponding approximately to the nominal layer thicknesses of 196 nm and 262 nm used in the optical simulations and discussion in the Main text.}
    \label{SI_AFM}
\end{figure}

Atomic force microscopy (AFM) images corresponding to the NPL films used in the reflectance measurements shown in Fig. 1 are presented in Fig. \ref{SI_AFM}. These images confirm the uniformity of the spin-coated NPL layers and provide an independent determination of the film thicknesses employed in the optical studies. The measured thicknesses corroborate the values used in the EM simulations and validate that the observed thickness-dependent reflectance modulation arises from controlled variations in the NPL layer thickness rather than from morphological inhomogeneity.

\section{Photoluminescence from NPLs on DBR}

\begin{figure}[h]
    \centering		
    \includegraphics[width=\textwidth]{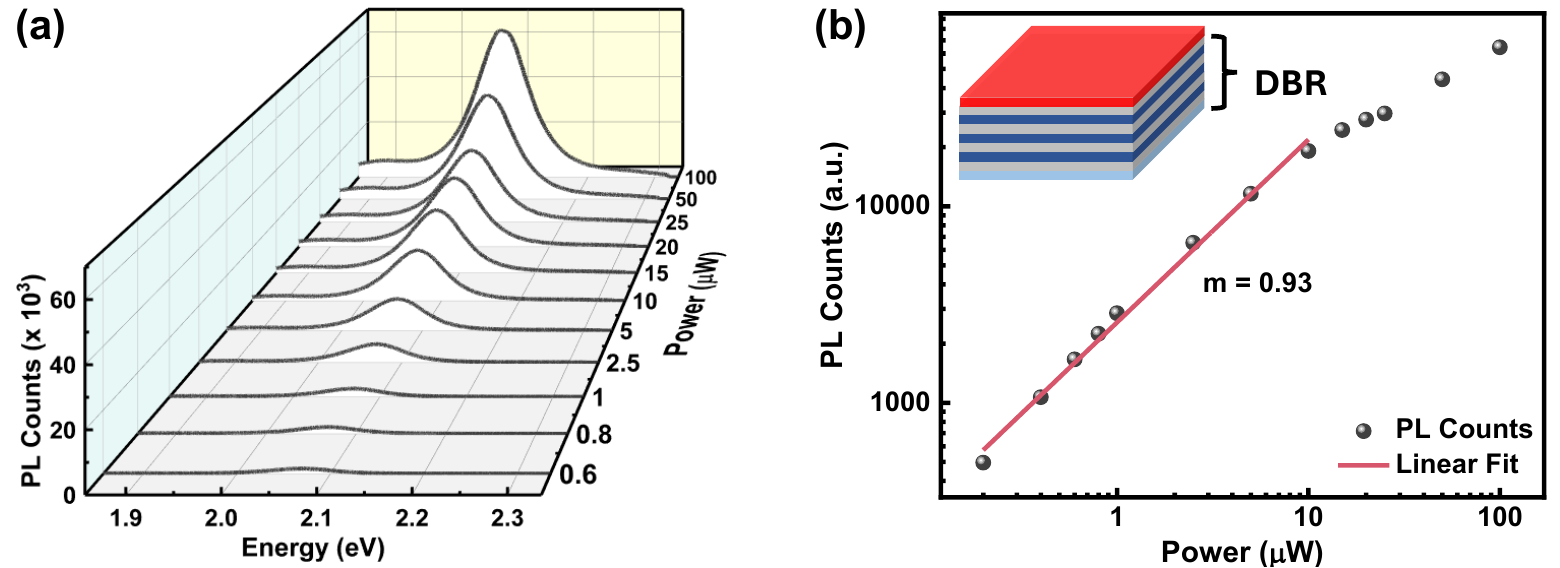}	 
    \caption{\textbf{(a)} Power-dependent PL spectra of C/S NPLs deposited on a DBR. \textbf{(b)} PL intensity as a function of excitation power on a log--log scale. The inset shows a schematic of the NPL layer deposited on the DBR.}    
    \label{SI_PL_NPL_DBR}
\end{figure}

To investigate whether the enhanced optical field on the DBR is sufficient to activate LH emission, power-dependent PL measurements were performed on a C/S-NPL film with a nominal thickness of 262 nm deposited on the DBR. A continuous-wave laser at 532 nm (2.33 eV) was used for excitation. As shown in Fig. \ref{SI_PL_NPL_DBR}, the integrated PL intensity increases approximately linearly with excitation power, with a power-law exponent of 0.93. Despite the enhanced PL intensity, no emission corresponding to the LH transition is observed. The DBR therefore primarily enhances the effective excitation and collection conditions without substantially modifying the intrinsic recombination pathway.

\section{Optical Response of Metallic Fabry--Perot Cavities}

\begin{figure}
    \centering
    \includegraphics[width=1\textwidth]{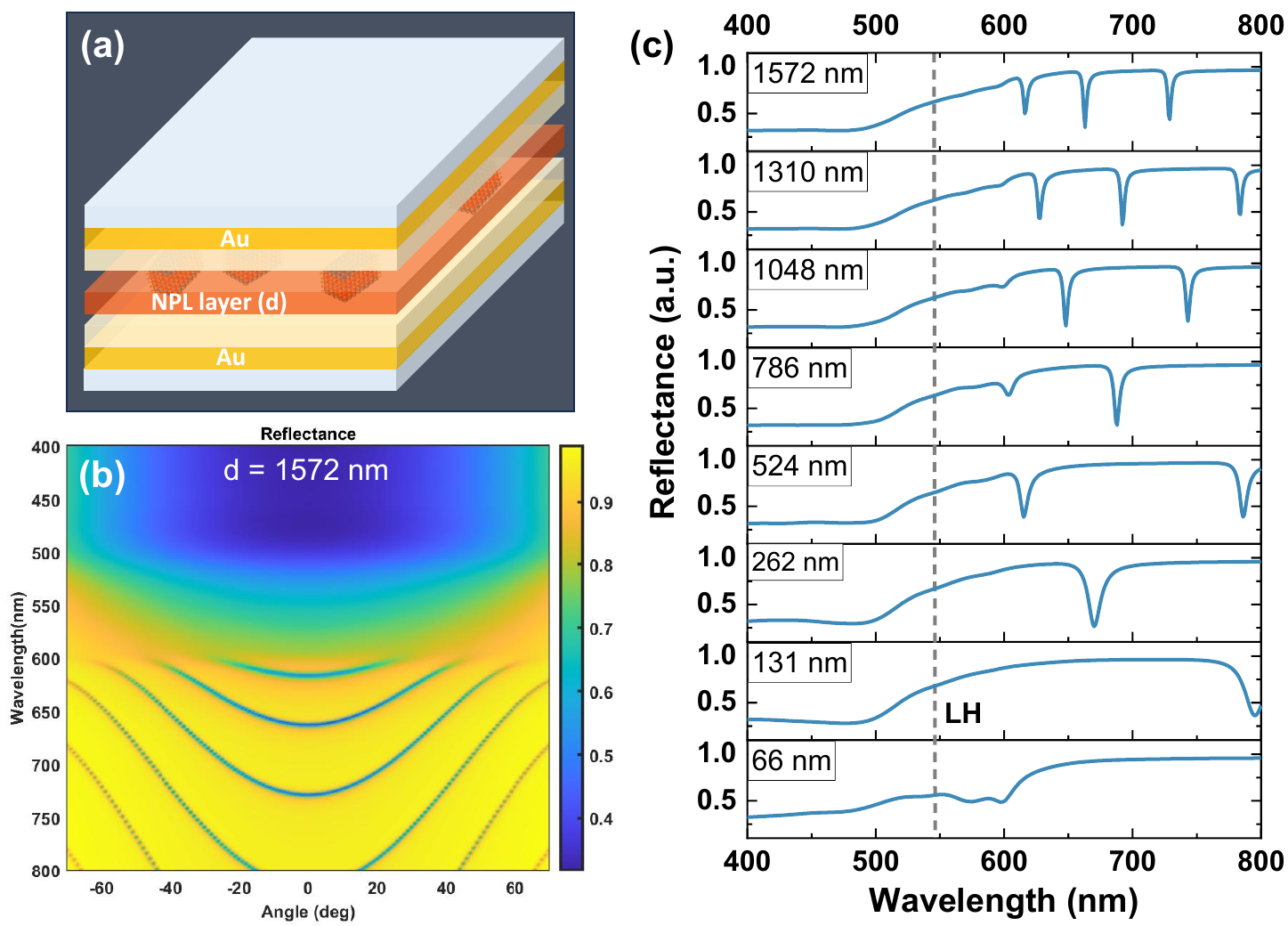}
    \caption{\textbf{(a)} Schematic of the Au–NPL–Au cavity with embedded NPL layer of thickness d. \textbf{(b)} Angle-resolved reflectance map for d = 1572 nm showing dispersive Fabry–Perot modes. \textbf{(c)} Normal-incidence reflectance spectra for varying cavity thicknesses (66–1572 nm).} 
    \label{SI_Au_cavity}
\end{figure}

Figure \ref{SI_Au_cavity}a shows the schematic representation of Au–NPL–Au Fabry–Perot cavity incorporating an NPL layer of thickness d. The optical response was calculated using the transfer matrix method, including the complex refractive index of Au to account for absorption losses.

The angle-resolved reflectance map for d = 1572 nm (Fig. \ref{SI_Au_cavity}b) reveals well-defined dispersive Fabry–Perot resonances whose spectral positions shift with angle as expected for a planar cavity. For wavelengths above 650 nm, the reflectance approaches $\simeq 95\%$, and relatively sharp cavity modes are observed. However, in the visible spectral range relevant to LH and HH excitons (500–640 nm), the reflectance decreases to 68\% due to increased absorption in the Au mirrors. 

Figure \ref{SI_Au_cavity}c shows normal-incidence reflectance spectra for varying cavity thicknesses (66–1572 nm). While spectrally narrow modes emerge for larger cavity thicknesses, their depth and contrast remain constrained by absorption-induced damping. For thinner cavities, resonances broaden and weaken further due to increased field penetration into the metallic layers.

These results indicate that although metallic Fabry–Perot cavities can support tunable and well-defined modes, their performance in the visible spectral range is fundamentally limited by mirror absorption. In systems such as colloidal NPLs or QDs, whose absorption and emission lie in the 500–640 nm range, this intrinsic loss reduces achievable quality factor, intracavity field buildup, and local density of optical states enhancement. In contrast, dielectric DBR cavities exhibit negligible absorption and high reflectivity across the stop band, thereby providing stronger and more spectrally selective optical confinement in this wavelength range.

\section{Exciton Decay Dynamics}

\begin{figure}
    \centering		\includegraphics[width=0.6\textwidth]{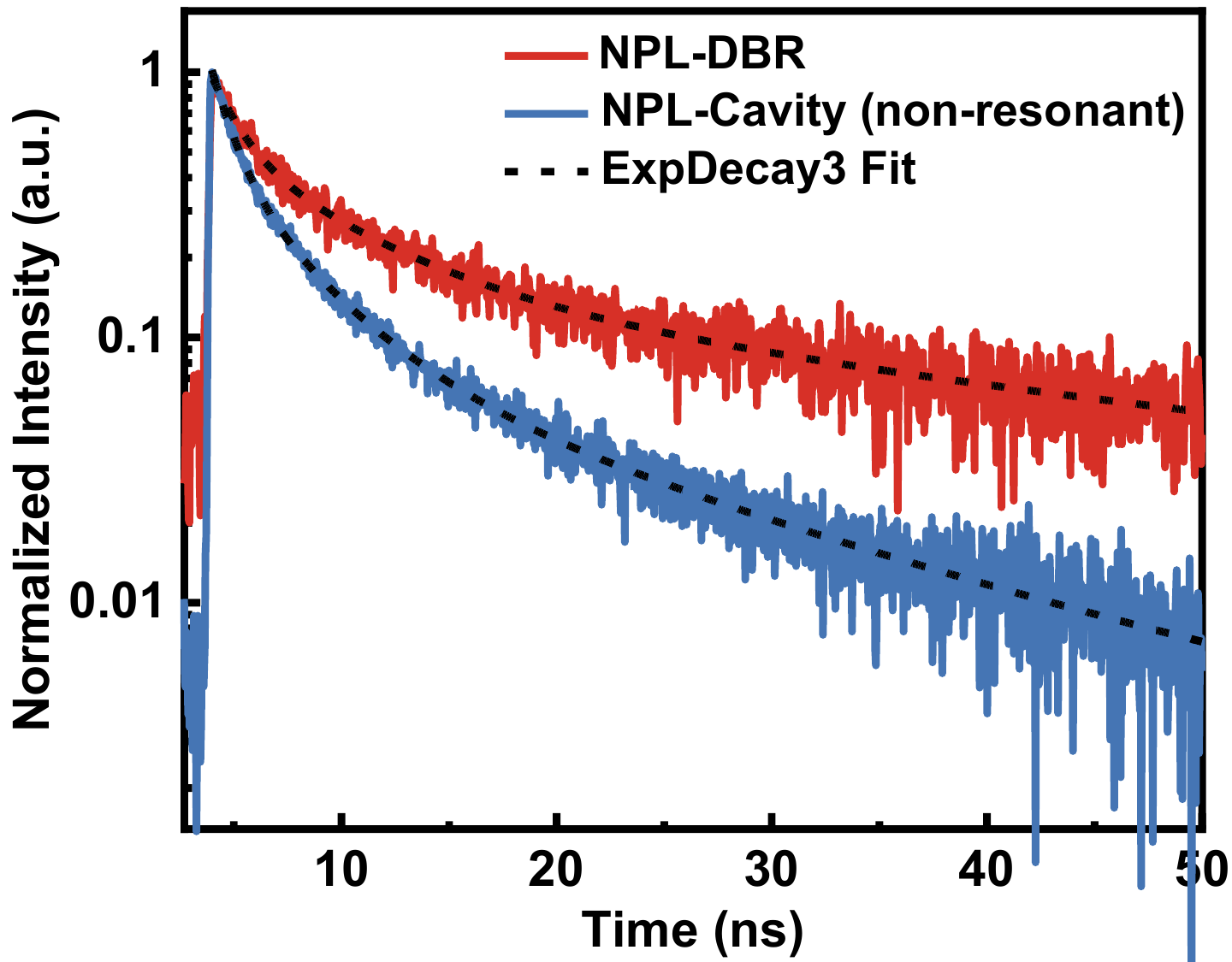}
    \caption{Exciton decay dynamics on DBR and inside non-resonant DBR cavity.} 
    \label{SI_NPL_TRPL}
\end{figure}

To further investigate the ASE behavior, we analyzed the exciton decay dynamics. Figure \ref{SI_NPL_TRPL} shows the comparison of exciton decay for NPLs on a DBR and NPLs inside the DBR cavity. The decay curves were fitted with a tri-exponential decay function given by,

\begin{equation}
    I(t)= \alpha_{1}e^{{- \frac{t}{\tau_{1}}}}+ \alpha_{2}e^{- \frac{t}{\tau_{2}}}+\alpha_{3}e^{- \frac{t}{\tau_{3}}}
    \label{eq:Tri-exp}
\end{equation}

and average lifetime was calculated using,

\begin{equation}
    \big<\tau\big> = \frac{\sum \alpha_{i}* \tau_{i}^{2}}{\sum \alpha_{i}*\tau_{i}}
    \label{eq:TRPL}
\end{equation}

The calculated average lifetimes are $22.95 \pm 1.21$ ns for NPLs on the DBR, and $7.84 \pm 0.47$ ns inside the non-resonant DBR cavity. The corresponding lifetime components and normalized weight factors are summarized in Table. \ref{tab:NPL_DBR_TRPL}.

\begin{table}
    \centering
    \begin{tabular}{ c  c  c  c  c  c  c  c } 
    \hline
    \hline
     System & $A_1$ & $\tau_1$ $(ns)$ & $A_2$ & $\tau_2$ $(ns)$ & $A_3$ & $\tau_3$ $(ns)$ & $\tau_{avg}$ $(ns)$ \\
    \hline
    \hline
    \\
    NPL-DBR & 0.47 & 1.30 & 0.38 & 5.77 & 0.15 & 33.04 & 22.95\\
     & $\pm$ 0.06 & $\pm$ 0.19 & $\pm$ 0.07 & $\pm$ 0.22 & $\pm$ 0.03 & $\pm$ 3.15 & $\pm$ 1.21\\ 
    \hline
    \\
    NPL-Cavity (non-resonant) & 0.58 & 0.98 & 0.33 & 3.78 & 0.08 & 16.14 & 7.84\\
    & $\pm$ 0.02 & $\pm$ 0.03 & $\pm$ 0.01 & $\pm$ 0.26 & $\pm$ 0.02 & $\pm$ 1.13 & $\pm$ 0.47\\
    \hline
    \\
    \end{tabular}
    \caption{\textbf{Summary of lifetime components along with weight factors for NPLs on DBR and NPL inside DBR cavity.}}
    \label{tab:NPL_DBR_TRPL}
\end{table}

The significant reduction in exciton lifetime in the cavity geometry is consistent with the observation of ASE. In the DBR cavity, the enhanced field strength and confinement effects lead to increased radiative recombination rates, resulting in shorter exciton lifetimes. This faster exciton decay supports the ASE action observed in the power-dependent PL measurements, as the higher radiative rates contribute to the conditions necessary for ASE \cite{zhang2020low,diroll2017violet}. 

\section{Temperature-Dependent PL of NPLs on Glass}

\begin{figure}
    \centering
    \includegraphics[width=1\textwidth]{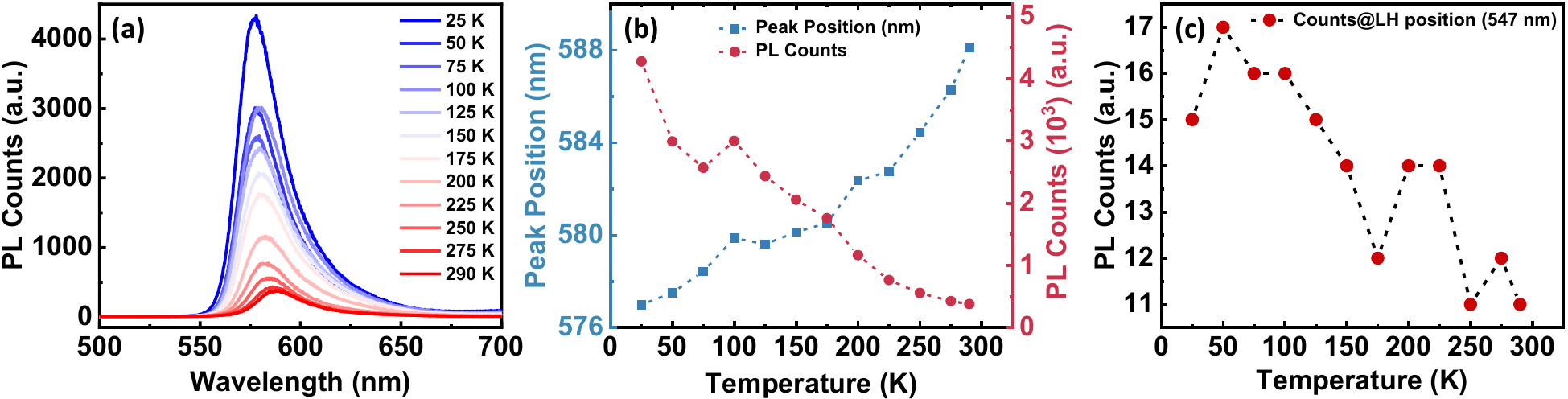}
    \caption{\textbf{(a)} Temperature dependent PL response from C/S NPLs deposited on glass substrate. \textbf{(b)} Variation of peak position and PL counts with temperature. \textbf{(c)} PL counts at the spectral position of LH (547 nm).} 
    \label{SI_NPL_Temp}
\end{figure}

Temperature-dependent PL measurements of NPLs on glass presented in Fig. \ref{SI_NPL_Temp} show the expected blue shift and increment in overall PL intensity with decreasing temperature, consistent with bandgap renormalization and reduced phonon-assisted non-radiative processes \cite{varshni1967temperature}. Notably, no distinct emission feature emerges at the LH spectral position across the full temperature range as shown in Fig. \ref{SI_NPL_Temp}c. The PL counts extracted specifically at the LH wavelength remain negligible, demonstrating that the LH exciton remains non-radiative even at low temperatures. This establishes that the absence of LH emission is intrinsic to the NPL electronic structure and not a consequence of thermal quenching.

\section{Excitation Power-Dependent PL}

\begin{figure}
    \centering
    \includegraphics[width=1\textwidth]{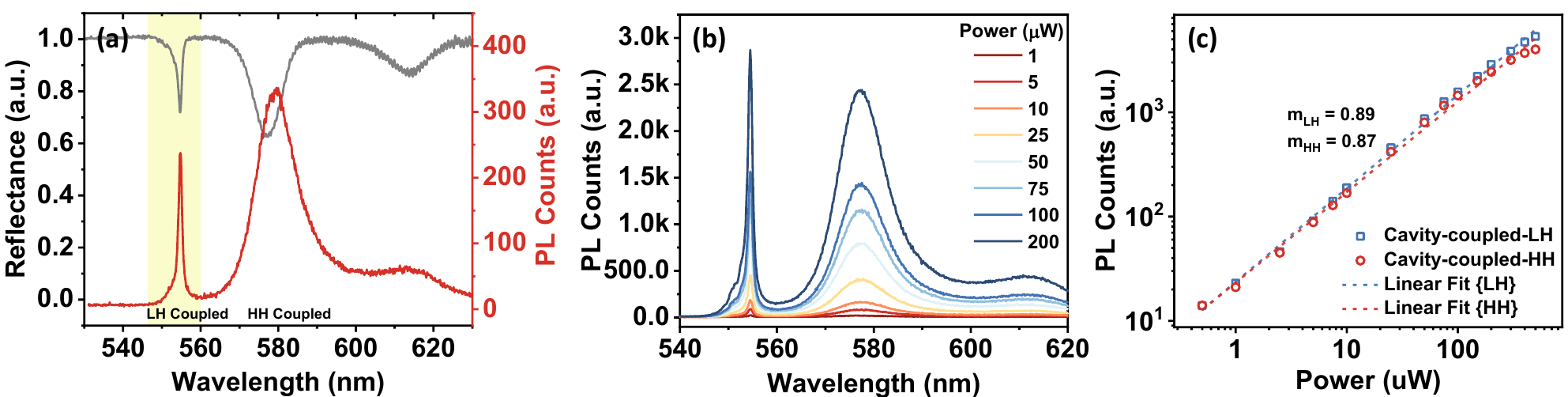}
    \caption{\textbf{(a)} Measured cavity reflectance (gray) at room temperature (297K) and the corresponding PL spectrum (red) from NPLs coupled to the cavity exhibits emission at the LH energy along with emission from HH coupled to the cavity. \textbf{(b)} PL spectrums measured at various incident laser power. \textbf{(c)} PL intensity as a function of pump power for the LH and HH-coupled to the cavity, plotted on a log–log scale.} 
    \label{SI_Power_Dep_LH}
\end{figure}

We also performed power-dependent PL measurements on the LH-coupled cavity as shown in Fig. \ref{SI_Power_Dep_LH}. The LH-related cavity-coupled feature and the HH emission both increase monotonically with excitation power without exhibiting any threshold-like behavior. Log-log analysis yields power-law exponents of $m_{LH} = 0.89$ and $m_{HH} = 0.87$, indicating that both features originate from spontaneous excitonic recombination rather than stimulated emission or ASE. A similar trend, in which HH- and LH-related nonlinear optical responses exhibit comparable pump-power dependences, has also been reported in pump-probe measurements of CdSe/CdS NPLs, suggesting comparable excitation-density scaling of the two excitonic populations under steady-state excitation.\cite{smirnov2019heavy}.

\section{Temperature-Dependent Reflectance of NPLs on DBR}

\begin{figure}
    \centering
    \includegraphics[width=0.5\textwidth]{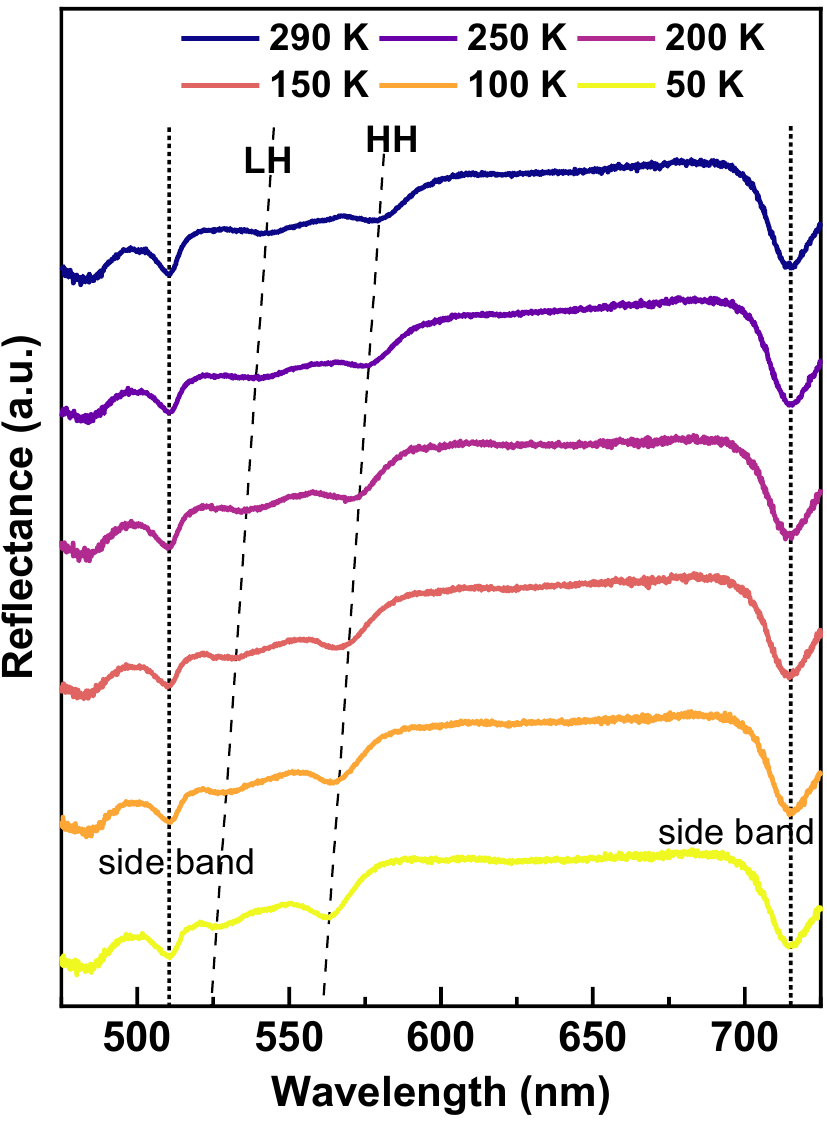}
    \caption{\textbf{(a)} Measured temperature dependent reflectance response from C/S NPLs on DBR exhibiting LH and HH blueshift with reducing temperature.} 
    \label{SI_Ref_Temp}
\end{figure}

Figure \ref{SI_Ref_Temp} shows the temperature-dependent white-light reflectance spectra of the C/S NPL layer deposited on the DBR substrate. As the temperature is reduced from 290 K to 50 K, both the LH and HH excitonic resonances exhibit a systematic blue-shift. A quantitative analysis of this blue-shift is already presented in Fig. 4(e) of the main. Both LH and HH transitions shift concurrently with temperature, leading to sequential spectral alignment with the cavity mode. As a result, the LH exciton comes into resonance with the cavity mode around room temperatures, while further cooling below  150 K brings the HH exciton into resonance. This temperature-driven tuning provides a controlled route to selectively access LH- and HH-dominated cavity coupling regimes.

\section{Temperature-Dependent Optical Response of Resonant Cavity}

\begin{figure}
    \centering
    \includegraphics[width=1\textwidth]{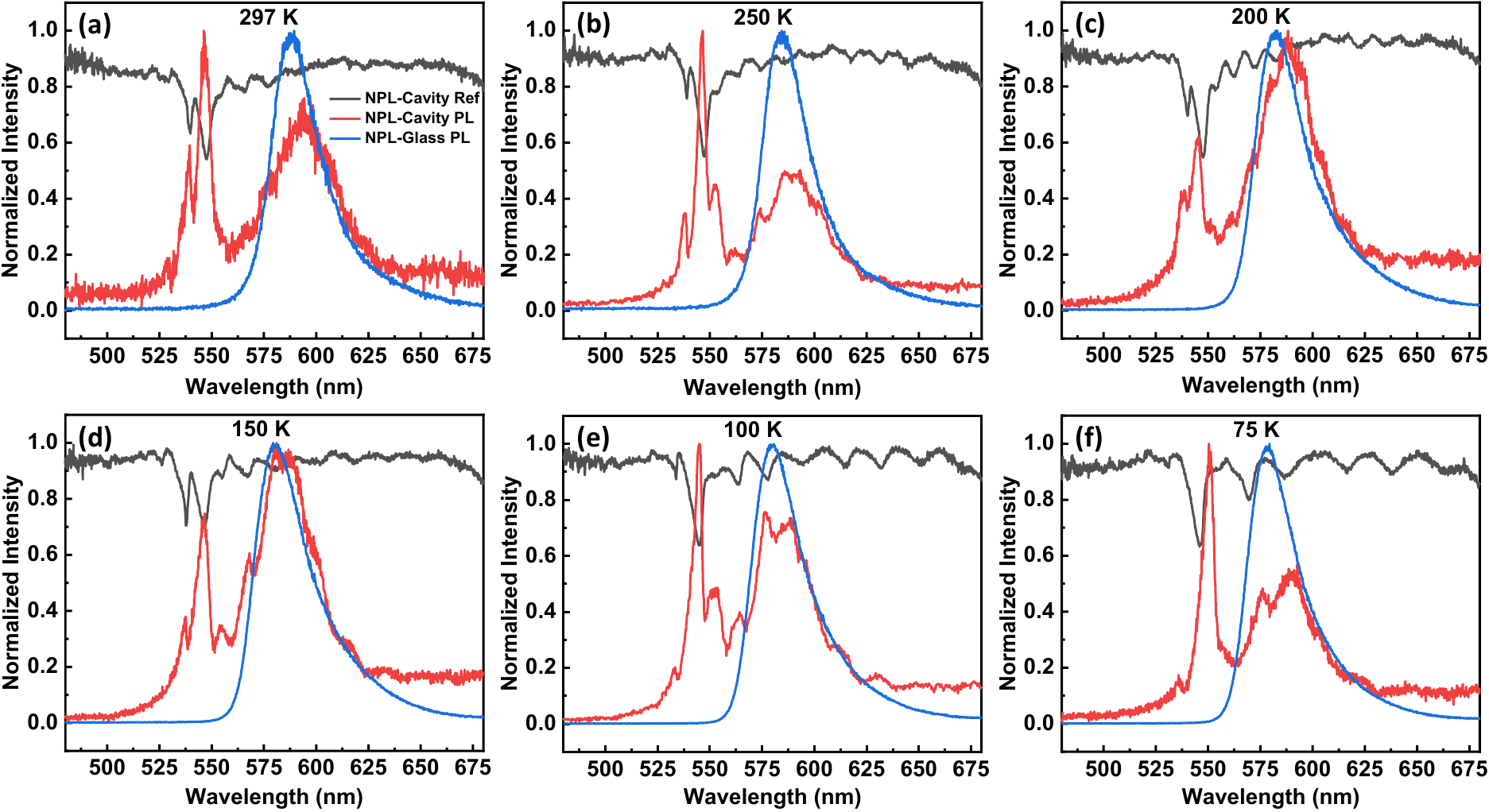}
    \caption{\textbf{(a)-(f)}  Measured cavity reflectance (gray) and PL (red) along with PL from NPLs on glass (blue) at various temperatures.} 
    \label{SI_NPL_DBR_LH}
\end{figure}

Figure \ref{SI_NPL_DBR_LH} shows the temperature-dependent cavity reflectance (gray), cavity-coupled PL emission (red), and PL emission from NPLs deposited on glass (blue). At all temperatures, the PL spectrum of NPLs on glass remains dominated by the HH exciton and shows no detectable emission at the LH spectral position, confirming that the LH exciton is intrinsically non-radiative in the absence of cavity coupling.

In contrast, the cavity-coupled PL exhibits enhanced emission at energies determined by spectral alignment between the cavity mode and the excitonic transitions. At higher temperatures, where the cavity mode is resonant with the LH exciton, a clear LH-related emission is observed in the cavity-coupled PL. Upon lowering the temperature, the concurrent blueshift of the LH and HH excitons leads to detuning of the LH exciton from the cavity mode, while further cooling brings the HH exciton into resonance and enhances HH-dominated cavity emission. These temperature-dependent trends are consistent with the behavior discussed in the Main text and further confirm that the observed emission enhancement arises from exciton–cavity spectral matching rather than intrinsic changes in the NPL emission properties.

\section{Optical Response of the HH-Control Cavity} 

\begin{figure}
    \centering
    \includegraphics[width=1\textwidth]{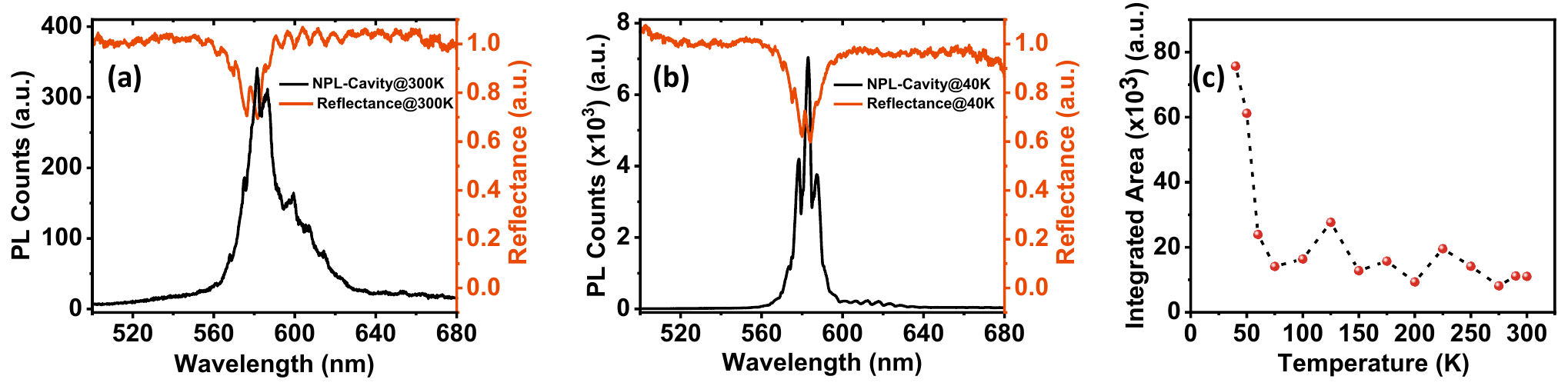}
    \caption{Measured cavity reflectance and PL spectra from the HH-control cavity at \textbf{(a)} 300 K, where the HH exciton is slightly detuned from the cavity mode, and \textbf{(b)} 40 K, where the HH exciton approaches resonance with the cavity mode. \textbf{(c)} Temperature dependence of the integrated cavity-coupled PL intensity, showing a monotonic increase as the HH exciton approaches spectral resonance.}
    \label{SI_Temp_PL_Cavity_HH}
\end{figure}

A separate cavity intentionally designed with a resonance close to the HH transition was investigated as a control. As the temperature is lowered, the HH exciton blueshifts toward the cavity resonance, producing a monotonic increase in the integrated cavity-coupled PL intensity, as shown in Fig. \ref{SI_Temp_PL_Cavity_HH}. This behavior differs from the non-monotonic response of the LH-coupled cavity and supports the interpretation that the temperature dependence is governed by exciton--cavity detuning.

\begin{figure}
    \centering
    \includegraphics[width=1\textwidth]{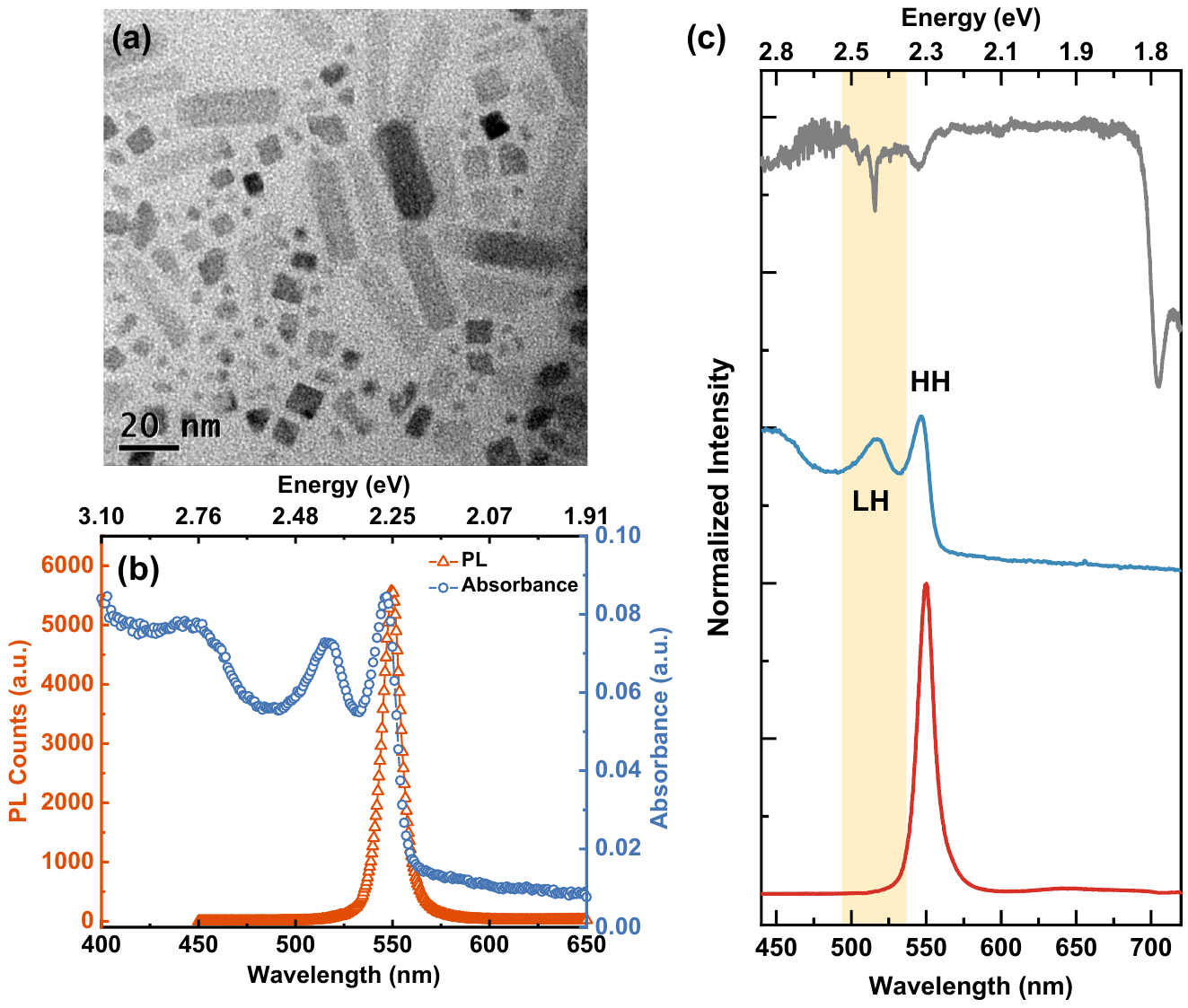}
    \caption{\textbf{(a)} TEM image of the synthesized C-NPLs. \textbf{(b)} PL emission and Absorbance of C-NPLs. \textbf{(c)} Measured cavity reflectance (gray) at room temperature (297 K) and NPL absorbance (blue), showing that the cavity mode is spectrally aligned with the LH absorption resonance. The corresponding cavity-coupled PL spectrum (red) shows no detectable emission at the LH energy, demonstrating that spectral resonance alone is insufficient and highlighting the importance of the material-dependent LH absorption strength.} 
    \label{SI_C_NPL_LH}
\end{figure}

\section{Core-Only NPLs in DBR Cavity}

Figure \ref{SI_C_NPL_LH} summarizes the structural and optical properties of C-NPLs and their response under cavity coupling. Panel (a) shows a TEM image confirming the successful synthesis and uniform morphology of the C-NPLs. Panel (b) presents the absorption and PL spectra, where the HH, LH, and SO bands are clearly visible in absorption, with PL dominated by the HH transition. Figure \ref{SI_C_NPL_LH}(c) shows the reflectance and PL spectra of C-NPLs embedded in a closed DBR cavity supporting a resonance near 518 nm, close to the LH absorption band. Although the cavity resonance is clearly observed in reflectance, no corresponding emission enhancement appears at the LH energy. The emergence of LH emission in the C/S NPL cavity, but not in the C-NPL cavity, indicates that cavity resonance must be accompanied by sufficient intrinsic LH absorption strength.

\section{Polarization-Resolved Measurements}

\begin{figure}
    \centering
    \includegraphics[width=1\textwidth]{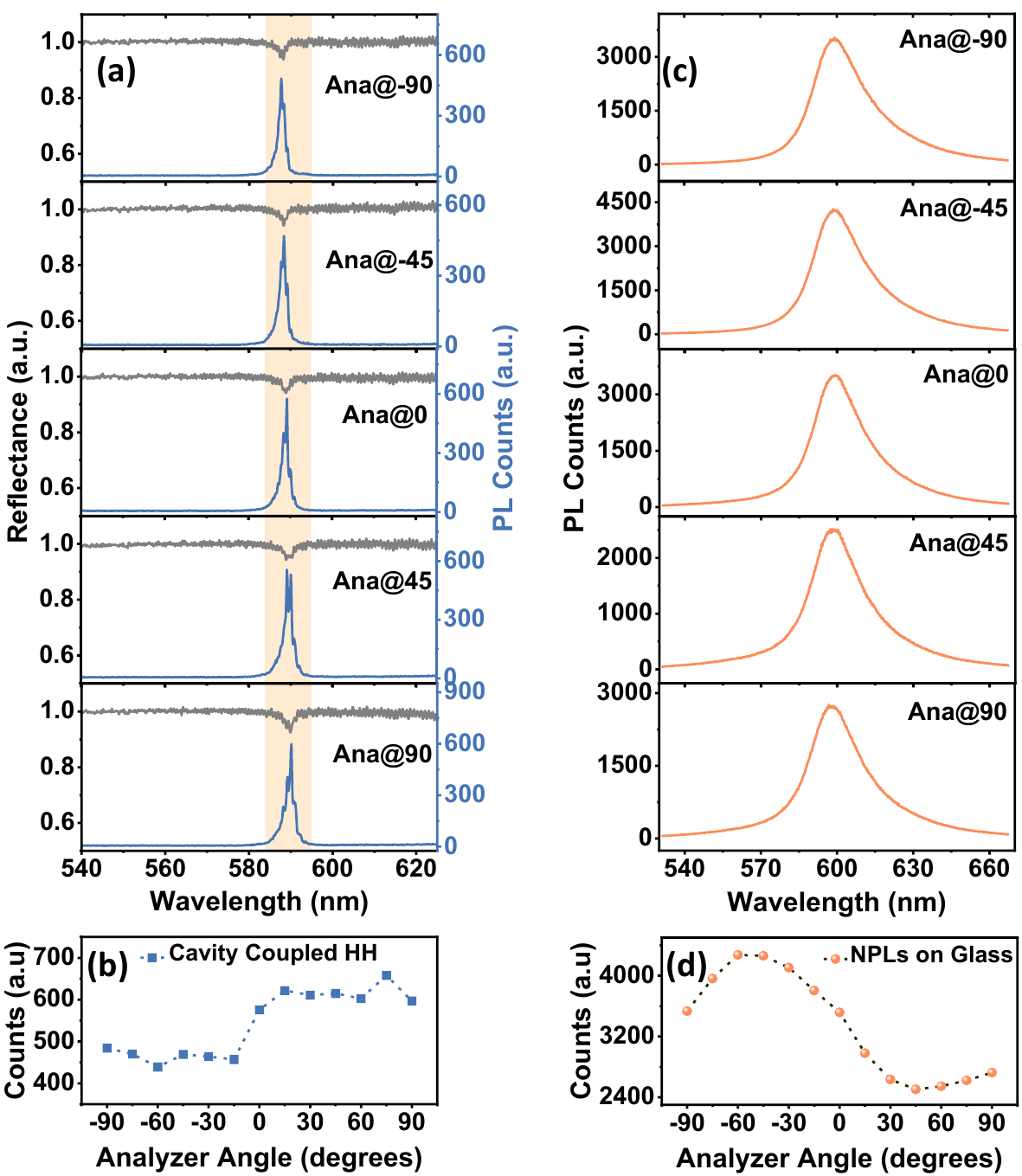}
    \caption{\textbf{(a)} Polarization-resolved cavity reflectance (gray) and PL spectra (red) measured from the cavity supporting resonance near the HH transition for analyzer angles ranging from $-90^\circ$ to $+90^\circ$, highlighting the cavity-coupled HH emission ($\sim$590 nm, shaded). \textbf{(b)} Peak PL intensity of the cavity-coupled HH emission as a function of analyzer angle. \textbf{(c)} Polarization-resolved PL spectra of uncoupled NPLs deposited on glass recorded over the same analyzer-angle range. \textbf{(d)} Peak PL intensity of the uncoupled HH emission from NPLs on glass as a function of analyzer angle.} 
    \label{SI_Cavity_Pol}
\end{figure}

As shown in Fig. \ref{SI_Cavity_Pol}(a,b), the cavity supporting resonance near the HH transition exhibits polarization-dependent modulation for the cavity-coupled HH ($\sim590$ nm) emission channel. For comparison, polarization-resolved PL measurements performed on uncoupled CdSe/CdS NPLs deposited on glass [Fig. \ref{SI_Cavity_Pol}(c,d)] provide the intrinsic polarization response of the NPL ensemble in the absence of cavity coupling. 
The corresponding cavity-coupled LH and HH polarization responses from the dual-mode cavity are presented in Fig. 5 of the Main manuscript. 

To quantitatively compare the polarization response, we calculated the degree of polarization (DOP) and the polarization extinction ratio (PER), defined as
\[
\mathrm{DOP}=\frac{I_{\mathrm{max}}-I_{\mathrm{min}}}{I_{\mathrm{max}}+I_{\mathrm{min}}},
\qquad
\mathrm{PER}=\frac{I_{\mathrm{max}}}{I_{\mathrm{min}}},
\]

where $I_{\mathrm{max}}$ and $I_{\mathrm{min}}$ are the maximum and minimum peak PL intensities extracted from the analyzer-angle-dependent measurements. The cavity-coupled LH emission exhibits the highest polarization modulation (DOP = 0.72), compared to the cavity-coupled HH emission in the dual-mode cavity (DOP = 0.28), the HH-control cavity (DOP = 0.20), and the uncoupled NPLs on glass (DOP = 0.26).

\begin{table}
\centering
\vspace{0.2cm}
\begin{tabular}{lcc}
\hline
\textbf{System} & \textbf{DOP} & \textbf{PER} \\
\hline
Cavity-coupled LH emission & 0.72 & 6.2 \\
Cavity-coupled HH emission (dual-mode cavity) & 0.28 & 1.7 \\
Cavity-coupled HH emission (HH-control cavity) & 0.20 & 1.5 \\
Uncoupled NPLs on glass & 0.26 & 1.7 \\
\hline
\end{tabular}
\caption{Degree of polarization (DOP) and polarization extinction ratio (PER) extracted from analyzer-angle-dependent peak PL intensity measurements.}
\label{tab:polarization}
\end{table}

\bibliography{References}

\end{document}